\documentclass[a4paper,11pt]{article}
\pdfoutput=1 

\usepackage[T1]{fontenc} 
\usepackage{amsmath,amssymb,mathtools}
\usepackage{color}
\usepackage{cite}
\usepackage[colorlinks=true,linkcolor=blueG, citecolor=redG, urlcolor=magentaG, bookmarks]{hyperref}
\usepackage{graphics}
\definecolor{comments}{RGB}{220,20,60}
\usepackage[text={17.6cm,24.6cm},centering]{geometry} 
\usepackage{braket}
\usepackage{comment}
\usepackage{tikz}

\usepackage{graphicx}
\usepackage{subcaption}
\usepackage{caption}
\usepackage{xcolor}
\usepackage{ifthen}

\newcommand{\nc}{\newcommand}
\nc{\ir}{\mathrm{i}}
\nc{\dd}{\mathrm{d}} 
\nc{\eE}{\mathrm{e}}
\nc{\Tr}{\text{Tr}}
\nc{\id}{\mathbb{I}}
\nc{\Z}{\mathcal{Z}}
\nc{\T}{\mathcal{T}}
\nc{\E}{\mathcal{E}}
\nc{\F}{\mathcal{F}}
\nc{\Om}{\Omega}
\nc{\N}{\mathcal{N}}
\nc{\spp}{\hspace{1pt}}
\nc{\spm}{\hspace{-1pt}}
\nc{\lb}{\label} 
\nc{\nn}{\nonumber} 
\nc{\ra}{\rangle} 
\nc{\la}{\langle} 
\nc{\Nn}{\mathcal{N}} 
\nc{\sigb}{\boldsymbol{\sigma_{\text{bd}}}}
\nc{\I}{\mathcal{I}}
\nc{\nnF}{\mathcal{F}}
\nc{\LE}{\mathcal{L}}
\nc{\AEc}{\mathcal{L}^\infty}
\nc{\M}{\mathcal{M}}

\definecolor{blueG}{RGB}{51, 102, 204}
\definecolor{magentaG}{RGB}{214.2, 40.8, 132.6}
\definecolor{redG}{RGB}{229.5, 51., 102.}

\usepackage{authblk}

\begin{document}
\title{\bf Reduced fidelities for free fermions out of equilibrium:\\ From dynamical quantum phase transitions to Mpemba effect}
 
\author[1]{Gilles Parez}
\affil[1]{ \it Laboratoire d'Annecy de Physique Th\'eorique (LAPTh), CNRS,\newline
Universit\'e Savoie Mont Blanc, 74940 Annecy, France}

\author[2]{Vincenzo Alba}
\affil[2]{\it Dipartimento di Fisica, Universit\`a di Pisa, and INFN Sezione di Pisa, \newline Largo Bruno Pontecorvo 3, Pisa, Italy}

\maketitle

\begin{abstract}
  We investigate the out-of-equilibrium dynamics after a quantum quench of the reduced fidelities
  between the states of a subregion $A$ at different times. Precisely, we consider the fidelity between the time-dependent state of $A$ and its initial value, as well as with the state at infinite time. We denote these fidelities as the reduced Loschmidt echo (RLE) and the final-state fidelity (FSF), respectively. If region $A$ is the full system, the RLE coincides with the standard Loschmidt echo. We focus on quenches from Gaussian states in several instances of the XY spin chain. 
  In the hydrodynamic limit of long times and large sizes of $A$, with their ratio fixed, the reduced fidelities admit a quasiparticle picture interpretation. Interestingly, for some quenches in the hydrodynamic regime the RLE features a complicated structure with an infinite sequence of nested lightcones, corresponding to quasiparticles with arbitrary large group velocities. This leads to a ``staircase'' of cusp-like singularities in the time-derivative of the fidelity. At the sub-hydrodynamic regime for some quenches the RLE exhibits cusp-like singularities, similar to the so-called dynamical quantum phase transitions (DQPT). We conjecture a criterion for the occurrence of the DQPT and for the ``critical'' times at which the singularities occur. Finally, we discuss the hydrodynamic limit of the FSF. In particular, we show that it provides a valuable tool to detect the so-called quantum Mpemba effect. 
  
\end{abstract}

\paragraph{Keywords:} Loschmidt echo, dynamical quantum phase transitions, quantum Mpemba effect
\newpage

\tableofcontents

\section{Introduction}

Quantum systems out of equilibrium host a variety of intricate emergent phenomena, such as dynamical phase transitions \cite{heyl2013dynamical}, entanglement spreading \cite{calabrese2005evolution,alba2017entanglement} or the quantum Mpemba effect \cite{ares2023entanglement,ares2025quantum}. Understanding nonequilibrium quantum matter is a fundamental challenge both in theoretical and experimental physics, with strong implications for the development of quantum technologies. Over the last two decades, there has been an important cross-fertilization between quantum information science and condensed matter physics. In particular, quantities originally studied in the realm of quantum information, such as entanglement measures \cite{VPRK97} and quantum fidelities \cite{U76,J94}, are now routinely used to probe quantum many-body properties such as quantum phase transitions \cite{OAFF02,ON02,VLRK03,CC04,zanardi2006ground,cozzini2007quantum,ZB08,parez2022symmetry}, topological phases of matter \cite{kitaev2006topological,levin2006detecting,hamma2008entanglement,abasto2008fidelity,eriksson2009reduced} and nonequilibrium dynamics of isolated systems \cite{calabrese2005evolution,alba2017entanglement,QSLZS06}. 

The simplest protocol to drive a quantum system out of equilibrium is a quantum quench \cite{calabrese2005evolution,cc-07}, where the system is initialized in a low-entangled state $|\Psi_0\rangle$ and evolves unitarily under the action of a Hamiltonian~$H$, such that the time-dependent state of the system is $|\Psi(t)\rangle = \eE^{-\ir t H}|\Psi_0\rangle.$ Typically the initial state $|\Psi_0\rangle$ is the ground state of some other Hamiltonian $H_0$ which does not commute with $H$, such that the quench is interpreted as an abrupt change in the system's Hamiltonian. For one-dimensional quantum integrable systems, the dynamics of entanglement and information after a quench is governed by the quasiparticle picture (QPP) \cite{calabrese2005evolution,fc-08,alba2017entanglement}. In its simplest form, this hydrodynamic approach proposes that entanglement spreads through the ballistic propagation of pairs of entangled quasiparticles emitted just after the quench. Quasiparticles emitted from different points in space are uncorrelated, whereas those emitted from the same location remain entangled through their propagation. In particular, QPP predicts the linear growth and saturation of the entanglement entropy after a quench, which has been observed experimentally \cite{kaufman-2016}. The quasiparticle picture also describes the propagation of other entanglement measures, and holds in multipartite settings \cite{fagotti2010entanglement,ac2-19, pbc-22,mac-21,parez2022analytical}. Recently, this powerful method has also been studied in two-dimensional systems \cite{gibbins2024quench,yamashika2024time} and for initial states with multiplets of entangled quasiparticles \cite{bastianello2018spreading,caceffo2023negative}, hence going beyond the pair paradigm. For free-fermionic systems, it is possible to obtain analytical results for the entanglement dynamics at all times \cite{fc-08,parez2022analytical,caceffo2023entanglement,alba2023logarithmic}.

In this paper, we investigate the dynamics of quantum fidelities in many-body systems out of equilibrium. A prominent example of nonequilibrium fidelity is the Loschmidt echo (LE) $\mathcal{L}(t)$, defined as the probability that the time-evolved system returns to its original state at some time $t$ \cite{peres1984stability},
\begin{equation}\label{eq:LE}
    \mathcal{L}(t) = |\langle \Psi_0|\Psi(t)\rangle|^2.
\end{equation}
For a system with $N$ degrees of freedom, the logarithmic LE, or rate function $\lambda(t)$, is then defined as 
\begin{equation}
\label{eq:rate}
    \lambda(t) = - \frac 1N \log(  \mathcal{L}(t)) .
\end{equation}
There is a rich literature which investigates the interplay between LE and quantum criticality. In Ref.~\cite{QSLZS06}, the authors study the dynamics of LE and observe that (i) it decays faster close to a critical point, and (ii) there are revivals whose frequencies depend on the system size. They suggest that there is an interesting relation between entanglement, LE and criticality, supported by \cite{happola2012universality}. Decay and revival of LE close to criticality has been observed in a variety of settings \cite{yuan2007loschmidt,venuti2010unitary,venuti2011exact,happola2012universality,montes2012phase,wang2015decay,lacki2019dynamical,hwang2019universality,jaiswal2022complexity}. However, there are counter examples \cite{jafari2017loschmidt}, namely criticality is not a sufficient nor necessary condition for enhanced decay and revival of LE. 

Going beyond the revival probability, the analytic structure of the rate function~\eqref{eq:rate} also contains valuable insights in the many-body dynamics. Indeed, for certain quantum systems in the thermodynamic limit $N\to \infty$, the rate function exhibits non-analyticities as a function of time. 
These phenomena are known as dynamical quantum phase transitions~(DQPT) \cite{heyl2013dynamical,heyl2014dynamical,heyl2018dynamical}. Generically, DQPTs are expected to occur when the initial state and the ground state of the quenched Hamiltonian belong to different quantum phases.
However, there are numerous exceptions to this behavior, and DQPT should be seen as a singular behavior distinct from equilibrium criticality \cite{heyl2018dynamical}.

Importantly, the LE defined in Eq.~\eqref{eq:LE} requires the knowledge of the time-evolved pure state of the full systems, which can be an experimental limitation. However, a recent experimental collaboration \cite{karch2025probing} introduced the subsystem Loschmidt echo (SLE), a quasi-local observable which exhibits the main features as the full LE, such as the DQPT, but which is experimentally accessible. In practice, the SLE is defined as products of projectors on the initial state averaged over various choices of subsystems.

When the whole system is in a pure state $|\Psi\rangle$, the state of a subsystem $A$ is described by the reduced density matrix (RDM) of $A$, defined as 
\begin{equation}
    \rho_A = \Tr_{\bar A} |\Psi\rangle \langle \Psi |.
\end{equation}
Here, $\bar A$ is the complement of $A$, such that $A \cup \bar A$ is the whole isolated system. In this paper, we focus on reduced fidelities $\mathcal{F}(\rho,\sigma)$, where both $\rho$ and $\sigma$ describe the same subsystem $A$ at different times.
Reduced fidelities have already been studied at equilibrium close to phase transitions \cite{parez2022symmetry}. Here we extend these investigations out of equilibrium. In particular, we introduce the reduced Loschmidt echo (RLE), defined as the fidelity between a subsystem RDM in its initial state, and at time $t>0$ after a quench. The RLE reduces to the SLE defined in \cite{karch2025probing} in the case where the initial state is a pure product state with translation invariance, but is otherwise more general. We focus on free-fermion systems, for which analytical calculations are possible, and numerical simulations can be performed for large subsystems. We show that the RLE exhibits DQPTs. Moreover, the RLE admits a QPP interpretation in the hydrodynamic limit. Interestingly, we show that for some initial states the RLE exhibits an intriguing kinematic structure featuring an infinite cascade of nested lightcones, similar to the full-counting statistics~\cite{groha2018full}.
We also define the final-state fidelity (FSF) as the fidelity between RDMs at time $t$ and in the stationary state for $t\to \infty$. This quantity provides a natural tool to probe late-time dynamics and thermalization in quantum many-body systems. We show that the FSF admits a much simpler QPP interpretation than the RLE. Finally, we show that the final state fidelity can be used as a tool to detect quantum Mpemba effects (or lack thereof). 

The paper is organized as follows. In Sec.~\ref{sec:fid} we introduce the various quantities of interest and express them in terms of two-point correlation functions in free-fermion systems. We discuss the quench protocols of interest in Sec.~\ref{sec:quench}, and we give the analytical time-dependent correlation functions in each case. In Secs.~\ref{sec:RLE} and \ref{sec:FSF} we provide our detailed analytical and numerical results for the RLE and the FSF, respectively. Finally, Sec.~\ref{sec:ccl} contains a summary of our main results and discusses future research directions.  

\section{Quantum fidelities}\label{sec:fid}

Here we introduce the quantum fidelities, which are the quantities of interest in this work. We provide the 
general definition of fidelity between two mixed-state systems in Sec.~\ref{sec:def-mix}. In Sec.~\ref{sec:FidelitiesGaussian} we outline the calculation of quantum fidelity for Gaussian fermionic states. 
We consider both fermionic states with well-defined number of fermions, as well as more generic states with fluctuating 
fermion number. In Sec.~\ref{sec:reduced} we introduce the reduced fidelities of interest. Precisely, we define the 
reduced Loschmidt echo and the final-state fidelity. 

\subsection{Definitions for mixed states}
\label{sec:def-mix}

A quantum fidelity quantifies how similar two states are. In the case of two (normalized) pure states $|\psi_\rho\rangle$, $|\psi_\sigma\rangle$, a fidelity $F$ can simply be defined as the squared overlap $F=|\langle \psi_\rho | \psi_\sigma \rangle |^2$. Clearly, the LE defined in Eq.~\eqref{eq:LE} is an example of pure-state fidelity. 
While pure states are ubiquitous, the most general quantum states are described by mixed states $\rho$ which satisfy the following physical conditions, (i) $\rho^\dagger = \rho$, (ii) $\rho \geqslant 0$ and (iii)~$\Tr \rho =1$. In the case of two mixed density matrices $\rho$, $\sigma$, the Uhlmann-Jozsa fidelity \cite{U76,J94} reads
\begin{equation}
    F_{\rm UJ}=\Tr \{(\sqrt{\rho}\sigma \sqrt{\rho})^{1/2}\}.
\end{equation}
This quantity belongs to the family of R\'enyi fidelities $\F_{m,n}$, defined as \cite{parez2022symmetry}
\begin{equation}
\F_{m, n}(\rho,\sigma) = \frac{\Tr\{(\rho^m \sigma^{2m}\rho^m)^n\}}{\sqrt{\Tr(\rho^{4nm})\Tr(\sigma^{4nm})}}. 
\end{equation}
Indeed, we have $\F_{1/2, 1/2} = F_{\rm UJ}$. In the following, we focus on the case $n=1$ and $m=1/2$, for simplicity, and drop the subscript. Our fidelity of interest is thus
\begin{equation}
\label{eq:fidelity}
\F(\rho,\sigma) = \frac{\Tr(\rho \sigma)}{\sqrt{\Tr(\rho^{2})\Tr(\sigma^{2})}},
\end{equation}
where we used the cyclicity of the trace. Importantly, $\F(\rho,\sigma)$ satisfies $0 \leqslant \F(\rho,\sigma) \leqslant 1$, where $\F(\rho,\sigma) = 1$ iff $\rho=\sigma$. Moreover, in the case of pure-state density matrices $\rho =| \psi_\rho\rangle \langle \psi_\rho |$ and $\sigma =| \psi_\sigma\rangle \langle \psi_\sigma |$, the fidelity reduces to the overlap, $\F(\rho,\sigma)= |\langle \psi_\rho | \psi_\sigma \rangle |^2$.

\subsection{Fidelities for Gaussian fermionic states}\label{sec:FidelitiesGaussian}

In this work we restrict ourselves to dynamics under quadratic fermionic Hamiltonians that preserve Gaussianity. This means that the if the initial state is Gaussian, the time dependent density matrix will be Gaussian at any time. 
In this section, we review how to express R\'enyi fidelities in terms of correlations matrices for Gaussian fermionic states. 

\paragraph{States that preserve the fermion number.} Let us introduce fermionic operators $c_j$ and $c_j^\dagger$, which satisfy the canonical anticommutation relations
\begin{equation}
\{c_j^\dagger, c_k\} = \delta_{jk}, \quad \{c_j, c_k\} =\{c_j^\dagger, c_k^\dagger\}=0.  
\end{equation}
For Gaussian states $\rho$ and $\sigma$  preserving the fermion number, namely 
\begin{equation}
\Tr(\rho c_j c_k) = 0 = \Tr(\rho c_j^\dagger c_k^\dagger), 
\end{equation}
and similarly for $\sigma$, we can express the fidelity $\F$ in terms of the correlations matrices $C_\rho$ and $C_\sigma$, defined as
\begin{equation}
[C_{\rho}]_{j,k} = \Tr(\rho c_j^\dagger c_k), \quad [C_{\sigma}]_{j,k} = \Tr(\sigma c_j^\dagger c_k).
\end{equation}
For simplicity we introduce the matrices 
\begin{equation}\label{eq:J}
J_{\rho} = 2C_{\rho}-\mathbb{I}, \quad J_{\sigma} = 2C_{\sigma}-\mathbb{I},
\end{equation}
and using algebra of Gaussian operators \cite{chung2001density,peschel2003calculation,peschel2009reduced,fagotti2010entanglement}, we have 
\begin{equation}\label{eq:TrRSJ}
\Tr(\rho \sigma) = \det \left(\frac{\mathbb{I}+J_\rho J_\sigma}{2}\right). 
\end{equation}
The fidelity defined in Eq.~\eqref{eq:fidelity} thus reads \cite{parez2022symmetry}
\begin{equation}\label{eq:TrFJ}
\F(\rho,\sigma) = \frac{\det(\mathbb{I}+J_\rho J_\sigma)}{\sqrt{\det(\mathbb{I}+J_\rho^2)\det(\mathbb{I}+ J_\sigma^2)}}.
\end{equation}

\paragraph{States that do not preserve the fermion number.} Let us now consider the case where $\rho$ and $\sigma$ are Gaussian states that do not preserve the fermion number. For instance, this happens for states of the $XY$ chain. 
For such states, we express the fidelity in terms of the covariance matrices $\Gamma_\rho$ and $\Gamma_\sigma$. To define them, we introduce Majorana operators $a_{2m-1}$ and $a_{2m}$, 
\begin{equation}
a_{2m-1} = c_m +c_m^\dagger, \qquad a_{2m} = \ir (c_m -c_m^\dagger),
\end{equation}
which satisfy the anticommutation relation $\{a_j, a_k\} = 2\delta_{j,k}$. The covariance matrix associated to the state~$\rho$ is 
\begin{equation}
\begin{pmatrix}\label{eq:cov}
 \Tr (\rho \ a_{2m-1}a_{2n-1} )& \Tr (\rho\   a_{2m-1}a_{2n})  \\  \Tr (\rho \  a_{2m} a_{2n-1}) &   \Tr (\rho \  a_{2m} a_{2n})
\end{pmatrix} =\delta_{m,n}\mathbb{I}_2+\ir[\Gamma_\rho]_{m,n}
\end{equation}
where $\mathbb{I}_2$ is the $2 \times 2$ identity matrix, and $[\Gamma_\rho]_{m,n}$ is thus a $2 \times 2$ block. Hence, the dimension of $\Gamma_\rho$ is twice that of the correlation matrix $C_\rho$. The definition of $\Gamma_\sigma$ is analogous. Thus, the generalization of Eq.~\eqref{eq:TrRSJ} reads \cite{fagotti2010entanglement,groha2018full}
\begin{equation}
\Tr(\rho \sigma) = \sqrt{\det \left(\frac{\mathbb{I}-\Gamma_\rho \Gamma_\sigma}{2}\right)}, 
\end{equation}
which is obtained by replacing $J_\rho\to\ir\Gamma_\rho$ in Eq.~\eqref{eq:TrRSJ}.
The fidelity is
\begin{equation}\label{eq:FGamma}
\F(\rho,\sigma) = \sqrt{\frac{\det(\mathbb{I}-\Gamma_\rho \Gamma_\sigma)}{\sqrt{\det(\mathbb{I}-\Gamma_\rho ^2)\det(\mathbb{I}- \Gamma_\sigma^2)}}}.
\end{equation}
Equation~\eqref{eq:FGamma} is the most general formula for fermionic Gaussian states, and it reduces to Eq.~\eqref{eq:TrFJ} for fermion number-preserving states, but in those cases the latter is simpler to manipulate. 

\subsection{Reduced fidelities and quantities of interest}
\label{sec:reduced}

In the following, we investigate reduced fidelities after a quantum quench, where the time-dependent state is $|\Psi(t)\rangle = \eE^{-\ir H t}|\Psi_0\rangle$. We consider a subsystem $A$, and the time-dependent RDM is thus $\rho_A (t)  = \Tr_{\bar A}|\Psi(t)\rangle \langle \Psi(t)|$.

\paragraph{Reduced Loschmidt echo.}The reduced Loschmidt echo (RLE) is the fidelity between RDMs at $t=0$ and $t>0$. We define it as
\begin{equation}\label{eq:RLE}
\F_A(t)  =  \frac{\Tr(\rho_A(0) \rho_A(t))}{\sqrt{\Tr(\rho_A(0)^{2})\Tr(\rho_A(t)^{2})}},
\end{equation}
which is a direct generalization of the usual pure-state LE defined in Eq.~\eqref{eq:LE}. In the case where the initial state is a translation-invariant product state, $|\Psi_0\rangle = |\psi\rangle \otimes \cdots \otimes |\psi\rangle$, the RLE reduces to the SLE introduced in Ref.~\cite{karch2025probing}. Otherwise, these are distinct quantities. For instance, the SLE requires the spatial average of products of local projectors onto the initial states over all subsystems, whereas the RLE considers a single well-defined subsystem $A$ and is defined for arbitrary initial states $|\Psi_0\rangle$, such as ground states of critical spin chains.

We also introduce the logarithmic RLE as
\begin{equation}\label{eq:lrLe}
\Lambda_A(t)= -\frac{1 }{\ell}\log( \F_A(t))
\end{equation}
where $\ell=|A|$ is the size of subsystem $A$. 

\paragraph{Final-state fidelity.} Our second quantity of interest is the final-state fidelity (FSF), which compares local states at time $t$ with the stationary state at $t\to \infty$. The whole isolated system remains in a pure state at all times, and hence never reaches a stationary state. However, for small enough subsystems $A$\footnote{More precisely, for finite subsystems $A$ embedded in infinitely-large systems.}, the complement $\bar A$ plays the role of the environment during the quench. As a result, local states typically relax to a stationary value \cite{calabrese2020entanglement}, 
\begin{equation}
    \rho_A^\infty = \lim_{t \to \infty}\rho_A(t).
\end{equation}
When this limit is well-defined, we introduce the FSF as
\begin{equation}\label{eq:FSF}
\F_A^\infty(t)  =  \frac{\Tr(\rho_A^\infty \rho_A(t))}{\sqrt{\Tr((\rho_A^\infty)^{2})\Tr(\rho_A(t)^{2})}},
\end{equation}
and the logarithmic FSF reads
\begin{equation}\label{eq:fsf}
\Lambda_A^\infty(t)= -\frac{1 }{\ell}\log (\F_A^\infty(t)).
\end{equation}

We stress that, as opposed to the RLE, the FSF does not admit a well-defined pure-states analog. Indeed, the natural idea would be to consider the following limit, 
\begin{equation}
 \lim_{t'\to \infty}\Big|\langle \Psi(t')| \Psi(t)\rangle\Big| =\lim_{t'\to \infty}\Big|\langle \Psi_0| \eE^{\ir H(t'-t)}|\Psi_0\rangle\Big|  =\lim_{t'\to \infty} \Big|\sum_n \eE^{\ir E_n (t'-t)}|\langle\Psi_0|\phi_n\rangle|^2\Big|,
\end{equation}
where $E_n$ and $|\phi_n\rangle$ are the eigenvalues and corresponding eigenstates of $H$. This limit contains highly-oscillating terms and is not well-defined.

\section{Model and initial states}\label{sec:quench}
In the rest of this paper, we consider the spin-$1/2$ XY spin chain in a transverse magnetic field with periodic boundary conditions. The Hamiltonian is 
\begin{equation}
\label{eq:HamiltonianXY}
H(h, \gamma)=-\sum_{j=1}^L \left( \frac{1+\gamma}{4}\sigma_{j}^x\sigma_{j+1}^x+\frac{1-\gamma}{4}\sigma_{j}^y\sigma_{j+1}^y +\frac{h}{2} \sigma_j^z \right),
\end{equation}
where $h$ is the external magnetic field, $\gamma$ is the anisotropy parameter, $L$ is the size of the system and $\sigma_j^{\alpha}$ are the Pauli matrices at site $j$. We impose periodic boundary conditions, such that $\sigma_{j+L}^{\alpha}\equiv \sigma_{j}^{\alpha}$. We systematically consider the case where subsystem $A$ consists of $\ell$ contiguous sites embedded in the periodic chain of length $L$ in the thermodynamic limit $L\to \infty$.

The XY Hamiltonian in Eq.~\eqref{eq:HamiltonianXY} can be mapped to a free-fermion model via the Jordan-Wigner transformation, and diagonalized by the Bogoliubov transformation \cite{lieb1961two}. For $\gamma=0$, the XY Hamiltonian reduces to the XX Hamiltonian. Importantly, the fermionic version of the XX model conserves the fermion number, but the generic XY model does not. Another special point is $\gamma=1$, where the XY model reduces to the quantum Ising chain. In the following, we investigate three distinct quench protocols. In all cases, the time-dependent RDMs are Gaussian, and the correlation functions are known in the thermodynamic limit $L\to \infty$ at all times.

\subsection{Quench from the N\'eel state in the XX chain}\label{sec:NeelQ}

In this quench protocol, we initialize the system in the N\'eel state. In terms of fermionic operators, it reads
\begin{equation}
    |\Psi_0\rangle = \prod_{j=1}^{L/2} c_{2j}^{\dagger} |0\rangle,
\end{equation}
where $ |0\rangle$ is the vacuum, which satisfies $ c_j|0\rangle=0$ for $j=1,2,\dots, L$. We evolve this state with the zero-field XX Hamiltonian, corresponding to $H(0,0)$ in Eq.~\eqref{eq:HamiltonianXY}. In the thermodynamic limit $L\to \infty$, the time-dependent two-point correlation matrix reads \cite{ac2-19}
\begin{equation}\label{eq:CANeel}
   [C_A]_{x,x'}= \langle \Psi(t)|c_x^\dagger c_{x'}|\Psi(t)\rangle = \frac{\delta_{x,x'}}2 +\frac{(-1)^{x'}}2\int_{-\pi}^{\pi}\frac{\dd k}{2\pi}\eE^{\ir k(x-x')+2\ir t\cos (k)},  \quad x,x'=1,2,\dots,\ell.
\end{equation}
Following Eq.~\eqref{eq:J} we introduce $J_A = 2C_A-\mathbb{I}$, and we have
\begin{equation}\label{eq:JANeel}
    [J_A(t)]_{x,x'} = (-1)^{x'}\int_{-\pi}^{\pi}\frac{\dd k}{2\pi}\eE^{\ir k(x-x')+2\ir t\cos (k)}, \quad x,x'=1,2,\dots,\ell.
\end{equation}

The two limiting cases $t=0$ and $t\to \infty$ yield
\begin{equation}
    [J_A(0)]_{x,x'} = (-1)^{x'} \delta_{x,x'}
\end{equation}
and
\begin{equation}
    \lim_{t \to \infty}  [J_A(t)]_{x,x'} = 0. 
\end{equation}

\subsection{Quench from the dimer state in the XX chain}
\label{sec:dimer-quench}

The dimer state is 
\begin{equation}
    |\Psi_0\rangle = \prod_{j=1}^{L/2}\frac{c_{2j}^{\dagger}-c_{2j-1}^{\dagger}}{2}|0\rangle. 
\end{equation}
Similar to the N\'eel quench, we evolve the dimer state with the zero-field XX Hamiltonian. Since it conserves the fermion number, we just need the $\ell \times \ell$ correlation matrix $C_A$, or equivalently $J_A = 2C_A-\mathbb{I}$. The matrix has a block-Toeplitz structure, 
\begin{equation}\label{eq:JAdim}
    J_A(t) = \begin{pmatrix}
\pi_0(t)& \pi_1(t) & \cdots & \pi_{\ell/2-1}(t)\\
\pi_{-1} (t)& \pi_0(t) &  &  \vdots\\
\vdots & &\ddots &\vdots \\
\pi_{1-\ell/2} (t)&\cdots &\cdots &\pi_0(t)
\end{pmatrix}.
\end{equation}
In the thermodynamic limit $L\to \infty$, the blocks read \cite{fagotti2014conservation,PBC21bis}
\begin{equation}
    \pi_{m}(t) =  \ \begin{pmatrix}
-f_m(t)&-g_m(t)\\
 -g_{-m}(t)&f_m(t) \\
\end{pmatrix}
\end{equation}
with
\begin{equation}
\left\{
\begin{aligned}
f_m(t) &=  \int_{-\pi}^{\pi}\frac{\dd k}{2 \pi}\eE^{-2\ir m k} \sin (k) \sin(2 \cos(k) t), \\
g_m(t) &=  \int_{-\pi}^{\pi}\frac{\dd k}{2 \pi}\eE^{-2\ir m k} \eE^{-\ir k}(\cos (k) + \ir \sin (k) \cos(2 \cos(k) t)).
\end{aligned}
\right.
\end{equation}

In the initial state we have 
\begin{equation}
\left\{
\begin{aligned}
f_m(0) &=  0, \\
g_m(0) &= \delta_{m,0}, 
\end{aligned}
\right.
\end{equation}
and in the stationary state we find
\begin{equation}
\left\{
\begin{aligned}
\lim_{t \to \infty}f_m(t) &=  0, \\
\lim_{t \to \infty}g_m(t) &= \frac 12 \delta_{m,0}. 
\end{aligned}
\right.
\end{equation}

\subsection{Quench from the XY ground state}
\label{sec:xy-ground}

Finally, we consider the generic quench protocol where the initial state $|\Psi_0\rangle$ is the ground state of the XY Hamiltonian $H(h_0, \gamma_0)$ with arbitrary parameters $h_0, \gamma_0$, and evolves under the action of $H(h, \gamma)$ with new parameters $h,\gamma$. In this situation, the time-dependent RDM is still Gaussian, but it does not preserve the fermion number. We thus use the $2\ell \times 2\ell$ covariance matrix introduced in Eq.~\eqref{eq:cov}. As for the dimer quench, the matrix has a block-Toeplitz structure, 
\begin{equation}
\label{eq:Toeplitz}
\Gamma_A(t)=\begin{pmatrix}
\Pi_0(t)& \Pi_1(t) & \cdots & \Pi_{\ell-1}(t)\\
\Pi_{-1}(t) & \Pi_0(t) &  &  \vdots\\
\vdots & &\ddots &\vdots \\
\Pi_{1-\ell}(t) &\cdots &\cdots &\Pi_0(t)
\end{pmatrix}, 
\end{equation}
where 
\begin{equation}
    \Pi_j(t)= \begin{pmatrix} -F_j(t) & G_j(t)\\
-G_{-j}(t) & F_j(t)\end{pmatrix}.
\end{equation}
In the large-$L$ limit, the entries read \cite{fc-08}
\begin{equation}
\label{eq:FG}
\left\{
\begin{aligned}
  F_j(t)&=\ir \int_{-\pi}^{\pi} \frac{\dd k}{2 \pi} \eE^{-\ir k j} \sin \Delta_k \sin (2 \epsilon_k t),  \\
  G_j(t)&= \int_{-\pi}^{\pi} \frac{\dd k }{2 \pi} \eE^{-\ir k j}\eE^{- \ir \theta_k} \left(\cos \Delta_k+\ir \sin \Delta_k \cos (2 \epsilon_k t) \right),
\end{aligned}
\right.
\end{equation}
with
\begin{equation}
\label{eq:quenchparameters}
\begin{split}
    \epsilon_k^2&=(h-\cos (k))^2+\gamma^2 \sin^2(k), \qquad \epsilon_{0,k}^2=(h_0-\cos (k))^2+\gamma_0^2 \sin^2(k),\\
    \eE^{-\ir \theta_k}&=\frac{\cos (k)-h-\ir \gamma \sin (k)}{\epsilon_k},\\
    \cos \Delta_k &=\frac{hh_0-\cos (k)(h+h_0)+\cos^2(k)+\gamma \gamma_0\sin^2 (k)}{\epsilon_k\epsilon_{0,k}},\\
    \sin \Delta_k& = -\sin (k) \frac{\gamma h_0-\gamma_0 h-\cos (k)(\gamma-\gamma_0)}{\epsilon_k \epsilon_{0,k}}.
\end{split}
\end{equation}

The values in the initial state read
\begin{equation}
\label{eq:FG0}
\left\{
\begin{aligned}
  F_j(0)&=0,  \\
  G_j(0)&= \int_{-\pi}^{\pi} \frac{\dd k }{2 \pi} \eE^{-\ir k j}\eE^{- \ir \theta_k} \eE^{\ir \Delta_k},
\end{aligned}
\right.
\end{equation}
and in the stationary state we have 
\begin{equation}
\label{eq:FGinf}
\left\{
\begin{aligned}
 \lim_{t \to \infty} F_j(t)&=0,  \\
 \lim_{t \to \infty} G_j(t)&= \int_{-\pi}^{\pi} \frac{\dd k }{2 \pi} \eE^{-\ir k j}\eE^{- \ir \theta_k}\cos \Delta_k .
\end{aligned}
\right.
\end{equation}

\section{Reduced Loschmidt echo: Exact results in the (sub-)hydrodynamic regime}\label{sec:RLE}

In this section we investigate the reduced Loschmidt echo in the different quench protocols and different scaling regimes. Precisely, we derive the QPP result describing the dynamics of the RLE in the hydrodynamic limit. For the quench from the dimer state in the XX chain we show that the RLE features an intriguing nested lightcone structure. We also consider the sub-hydrodynamic regime, showing that the RLE exhibits singularities as a function of time, similar to the Loschmidt echo.

\subsection{Analytical results for the N\'eel quench: QPP and out-of-equilibrium singularities}
\label{sec:Neel}

We start by considering the quench from the N\'eel state, introduced in Sec.~\ref{sec:NeelQ}. In the definition of the RLE in Eq.~\eqref{eq:RLE}, the denominator is essentially a product of entanglement entropies, whose full quench dynamics is known analytically \cite{fc-08,PBC21bis}. We thus focus on the numerator, 
\begin{equation}
  \Tr(\rho_A(0) \rho_A(t)) = \det\left(\frac{\mathbb{I}+J_A(0)J_A(t)}{2}\right),
\end{equation}
where $J_A(t)$ is given in Eq.~\eqref{eq:JANeel}. Since our quantity of interest is the logarithmic RLE, we are interested in the quantity
\begin{equation}
    \log  \det\left(\frac{\mathbb{I}+J_A(0)J_A(t)}{2}\right) = \Tr \log \left(\frac{\mathbb{I}+J_A(0)J_A(t)}{2}\right).
\end{equation}
To proceed, we introduce $c_n$ as the Taylor coefficients of the function $\log((1+x)/2)$, 
\begin{equation}
    \log\Big(\frac{1+x}{2}\Big) = \sum_{n=0}^{\infty}c_n x^n,
\end{equation}
and we define the quantity $\M_n = \Tr(J_A(0)J_A(t))^n$. Combining these definitions, we have 
\begin{equation}
     \label{eq:series}
     \log \Tr(\rho_A(0) \rho_A(t)) = \sum_{n=0}^\infty c_n \M_n. 
\end{equation}
Our goal is to evaluate $\M_n$ analytically. This calculation is amenable using multidimensional stationary phase approximation \cite{fc-08}, which holds in the limit where $\ell$ and $t$ are large.  

For $n=0,1$, it is straightforward to evaluate $\M_n$. One obtains
\begin{equation}\label{eq:TrJJ01}
\begin{split}
\M_0 &= \ell, \\
\M_1 &= \ell \int_{-\pi}^\pi \frac{\dd k}{2 \pi} \eE^{2\ir t \cos (k)}.
\end{split}
\end{equation}

The calculation of $\M_2$ is more involved. The matrix elements of the product $J_A(0)J_A(t)$ are
\begin{equation}
[J_A(0)J_A(t)]_{x,x'} = (-1)^{x+x'} \int_{-\pi}^\pi \frac{\dd k}{2 \pi} \eE^{\ir k (x-x')+2\ir t \cos (k)}. 
\end{equation}
With the identity
\begin{equation}
\label{eq:sumexp}
    \sum_{m=1}^{\ell}\eE^{\ir k m}=\eE^{\ir k \left(\frac{\ell+1}{2} \right)}\frac{\sin\left(\frac{\ell}{2}k \right)}{\sin\left( \frac{k}{2}\right)},
\end{equation}
we find 
\begin{equation}
\M_2 = \int_{-\pi}^\pi \frac{\dd k_1}{2 \pi}\int_{-\pi}^\pi \frac{\dd k_2}{2 \pi}\left[\frac{\sin\left(\frac{\ell}{2}(k_1-k_2) \right)}{\sin\left(\frac{(k_1-k_2)}{2} \right)}\right]^2\eE^{2\ir t (\cos (k_1)+\cos (k_2))}. 
\end{equation}
Using standard techniques of multidimensional stationary phase approximation \cite{fc-08}, we arrive at 
\begin{equation}
\M_2 = \int_{2 |v_k| t < \ell} \frac{\dd k}{2 \pi} \left(\eE^{2\ir t \cos (k)}\right)^2 (\ell-2|v_k| t), 
\end{equation}
or equivalently
\begin{equation}
\M_2 = \ell \int_{-\pi}^\pi  \frac{\dd k}{2 \pi} \left(\eE^{2\ir t \cos (k)}\right)^2 -\int_{-\pi}^\pi  \frac{\dd k}{2 \pi} \left(\eE^{2\ir t \cos (k)}\right)^2 \min(2|v_k| t,\ell)
\end{equation}
where $v_k=\epsilon_k'= \sin (k)$ is the velocity of quasiparticles with momenta $k$. 

For $n \geqslant 2$, the calculations generalize directly and we have 
\begin{equation}\label{eq:TrJJn}
\M_n = \ell \int_{-\pi}^\pi  \frac{\dd k}{2 \pi} \left(\eE^{2\ir t \cos (k)}\right)^n -\int_{-\pi}^\pi  \frac{\dd k}{2 \pi} \left(\eE^{2\ir t \cos (k)}\right)^n \min(2|v_k| t,\ell). 
\end{equation}

Combining Eqs.~\eqref{eq:TrJJ01} and \eqref{eq:TrJJn}, we find 
\begin{equation}
\sum_{n=0}^\infty c_n \M_n = \ell \sum_{n=0}^\infty \int_{-\pi}^\pi  \frac{\dd k}{2 \pi} c_n \left(\eE^{2\ir t \cos (k)}\right)^n -\sum_{n=2}^\infty \int_{-\pi}^\pi  \frac{\dd k}{2 \pi} c_n \left(\eE^{2\ir t \cos (k)}\right)^n \min(2|v_k| t,\ell). 
\end{equation}
The first Taylor coefficients are 
\begin{equation}
c_0 = -\log (2), \qquad c_1=1,
\end{equation}
and therefore we have 
\begin{multline}\label{eq:numNeel}
\sum_{n=0}^\infty c_n  \M_n = \int_{-\pi}^\pi  \frac{\dd k}{2 \pi}  \log \left(\frac{1+\eE^{2 \ir t \cos (k)}}{2}\right)(\ell -\min(2|v_k| t,\ell))+ \int_{-\pi}^\pi  \frac{\dd k}{2 \pi} (-\log (2) +  \eE^{2 \ir t \cos (k)})\min(2|v_k| t,\ell).
\end{multline}
Notice that the oscillating terms in Eq.~\eqref{eq:numNeel} depend only on time $t$ and 
in the limit $t\to\infty$ give subleading contributions in the hydrodynamic limit. Indeed, these terms are usually neglected in the stationary phase approximation. Here we keep them because, as it will be clear in the following, they account for the singularities of the RLE. 

To compute the RLE, we also need the denominator in Eq.~\eqref{eq:RLE}. We have \cite{fc-08,PBC21bis}
\begin{equation}\label{eq:S2Neel}
    \log \Tr \rho_A(t)^2 = -\int_{-\pi}^\pi  \frac{\dd k}{2 \pi} \log (2) \min(2|v_k| t,\ell),
\end{equation}
and trivially $\log \Tr \rho_A(0)^2=0 $. Combining Eqs.~\eqref{eq:numNeel} and \eqref{eq:S2Neel} in Eq.~\eqref{eq:lrLe}, we find
\begin{multline}\label{eq:LRLENeel}
\Lambda_A(t) = \frac 12\int_{-\pi}^{\pi} \frac{\dd k}{2\pi} \log (2) \min(2|v_k| t/\ell,1)-\int_{-\pi}^{\pi} \frac{\dd k}{2\pi} \log \left(\frac{1+\eE^{2 \ir t \cos (k)}}{2}\right) (1-\min(2|v_k| t/\ell,1)) \\ - \int_{-\pi}^{\pi} \frac{\dd k}{2\pi}  \eE^{2 \ir t \cos (k)}\min(2|v_k| t/\ell,1).
\end{multline}
The terms $\min(2|v_k| t/\ell,1)$ in Eq.~\eqref{eq:LRLENeel} signal that the RLE is governed by the dynamics of quasiparticles with velocities $v_k$, similarly to the entanglement entropy and other entanglement measures \cite{calabrese2005evolution,fc-08,alba2017entanglement}. Again, in standard QPP calculations, all quantities depend only on the ratio $t/\ell$, such that we typically consider the hydrodynamic regime where both $t,\ell\to \infty$ with a fixed ratio $t/\ell=\textrm{cst}$. Here however, we have oscillatory terms of the form $\eE^{2 \ir t \cos (k)}$, which only depend on $t$. We thus consider two distinct regimes,  (i) a sub-hydrodynamic regime, where $0 \ll t \ll \ell $ and oscillations of the RLE are non-negligible, and (ii) the hydrodynamic regime where both $t$ and $\ell$ are infinite with a fixed ratio. In the hydrodynamic regime, the oscillatory terms vanish and the logarithmic RLE reads
\begin{equation}\label{eq:LRLENeelHydro}
\Lambda_A(t) =   \log (2)-\frac 12\int_{-\pi}^{\pi} \frac{\dd k}{2\pi} \log (2) \min(2|v_k| t/\ell,1) .
\end{equation}

\begin{figure}
    \centering
    \includegraphics[width=0.47\linewidth]{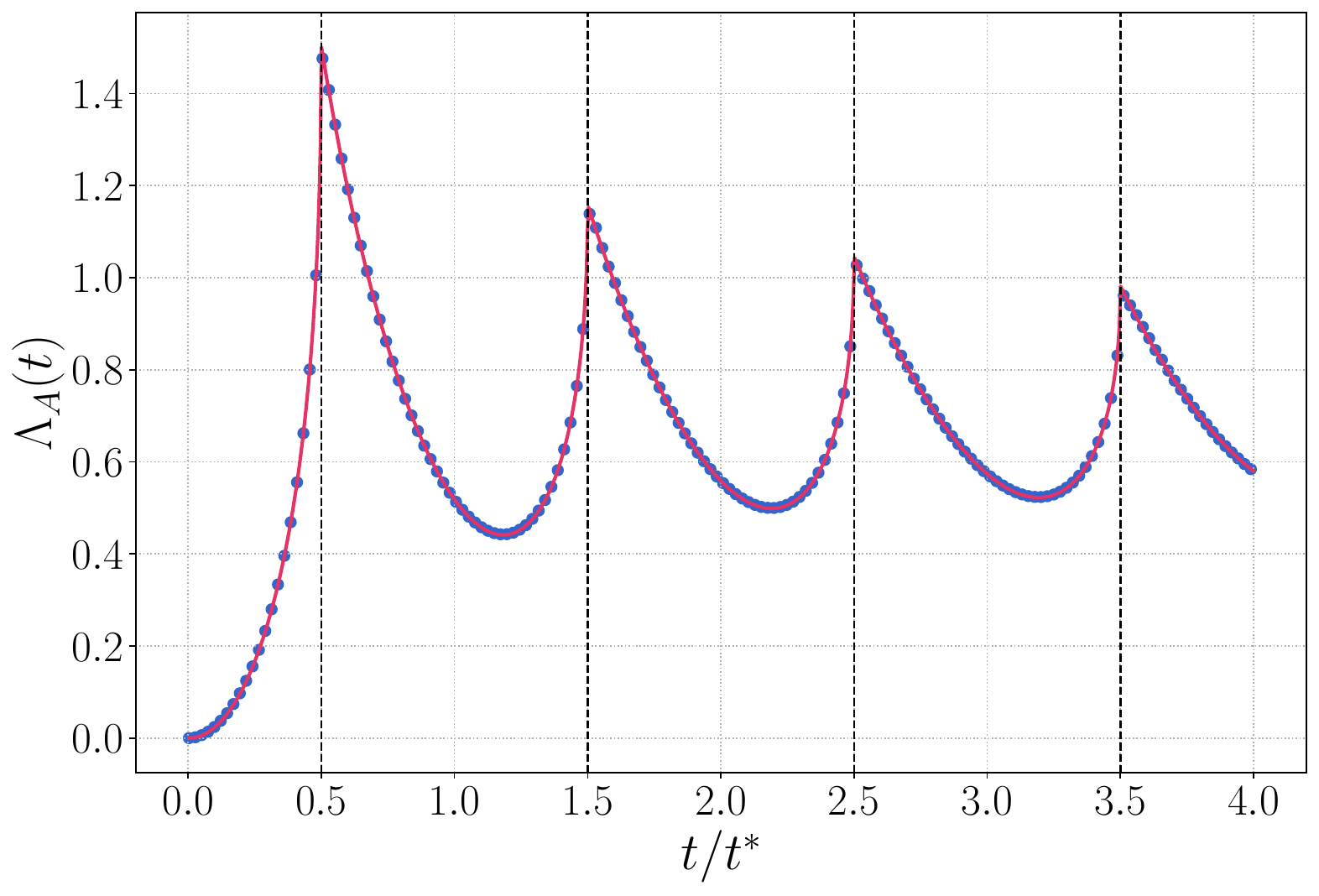}
     \includegraphics[width=0.47\linewidth]{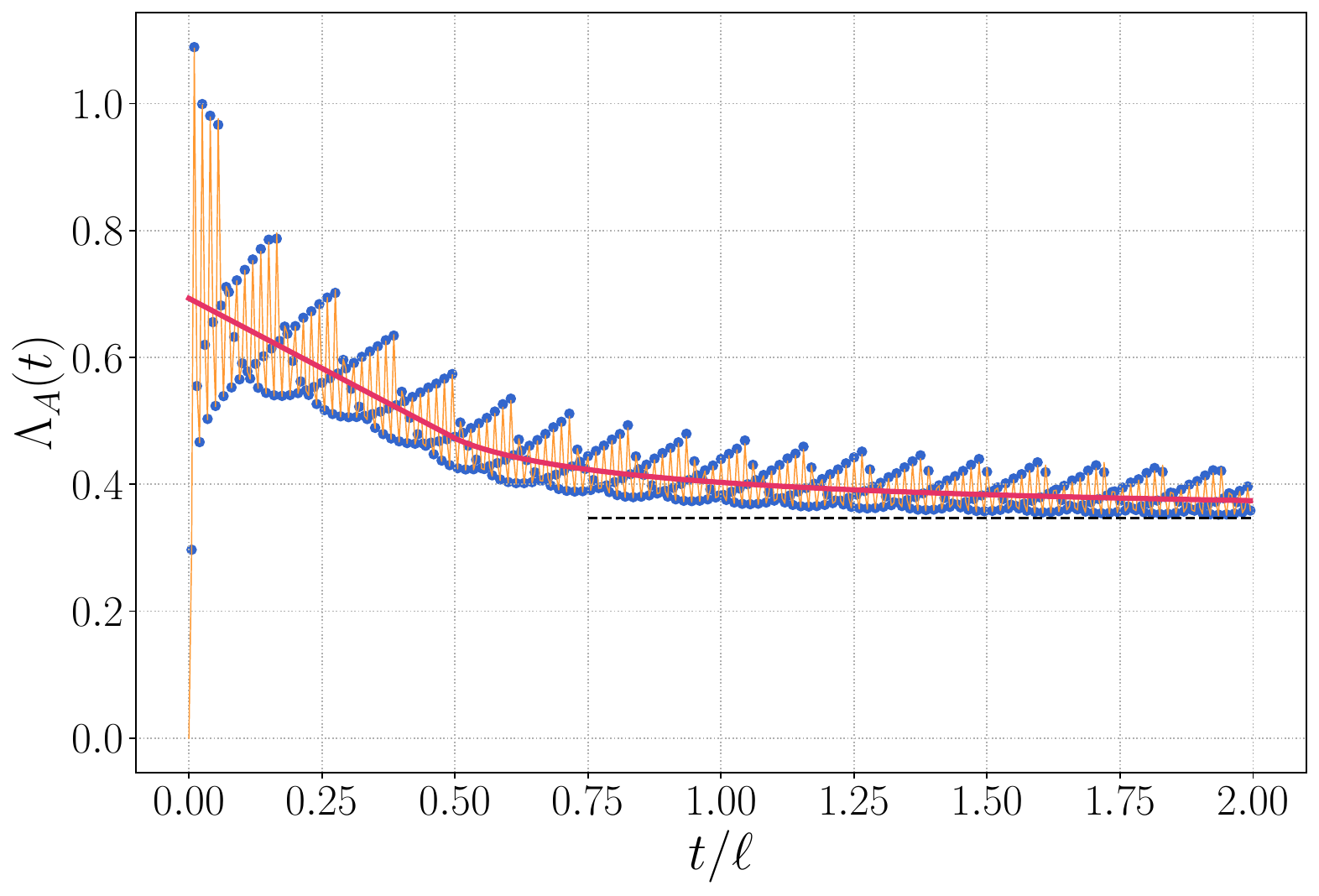}
    \caption{Logarithmic RLE in the XX chain from the N\'eel state, in both panels the symbols are obtained by exact numerical diagonalization with $\ell=200$. \textit{Left:} Logarithmic RLE as a function of $t/t^*$ with $t^* = \pi$. The solid line is the theoretical prediction of Eq.~\eqref{eq:LRLENeel} in the sub-hydrodynamic regime. We observe that DQPTs occur at $t=(m+1/2)t^*$ where $m$ is an integer. \textit{Right:} Logarithmic RLE as a function of $t/\ell$, in the hydrodynamic regime. The solid lines are theoretical predictions. The thin orange line is Eq~\eqref{eq:LRLENeel}, whereas the thick red one is Eq.~\eqref{eq:LRLENeelHydro}. Finally, the dotted horizontal line is the stationary value $1/2 \log(2)$. We observe that the oscillations are perfectly captured by Eq.~\eqref{eq:LRLENeel}, and that they are still strong at $\ell=200$, but decrease in amplitude with $t$.}
    \label{fig:RLENeel}
\end{figure}

We observe that Eq.~\eqref{eq:LRLENeel} exhibits non-analyticities due to the term $\log \big(1+\eE^{2 \ir t \cos (k)}\big)$, and more specifically its time-derivative, 
\begin{equation}\label{eq:mukNeel}
    \frac{\dd}{\dd t}\log \big(1+\eE^{2 \ir t \cos (k)}\big) = 2\ir \cos(k) \frac{\eE^{2 \ir t \cos (k)}}{1+\eE^{2 \ir t \cos (k)}}. 
\end{equation}
This term is singular whenever $\eE^{2 \ir t \cos (k)}=-1$, but contributions from $k$ and $(\pi-k)$ cancel in the integral, except for $k=0$. Therefore, Eq.~\eqref{eq:LRLENeel} is non-analytic for $t=(m+1/2)t^*$, where $m$ is an integer and $t^*=\pi$. Hence, DQPTs occur at times governed by the critical time $t^*=\pi$, which corresponds to the one obtained for the full LE in the same quench \cite{andraschko2014dynamical}. However, we stress that the RLE detects DQPTs only in the sub-hydrodynamic regime, i.e., at relatively short times. For $t$ very large, the oscillatory terms become negligible and we enter the hydrodynamic regime, where the RLE is analytical at all times. Finally, the RLE reaches a stationary value, $\lim_{t\to \infty}\Lambda_A(t) = 1/2 \log(2)$. In Fig.~\ref{fig:RLENeel} we compare our predictions of Eqs.~\eqref{eq:LRLENeel} and \eqref{eq:LRLENeelHydro} with exact numerical diagonalization based on the free-fermion method discussed in Sec.~\ref{sec:FidelitiesGaussian} and the correlation matrix given in Eq.~\eqref{eq:CANeel}. We find an excellent agreement between our analytical predictions and numerical results. In particular, (i) the oscillations in the logarithmic RLE are perfectly captured by Eq.~\eqref{eq:LRLENeel}, and (ii) the logarithmic RLE approaches the stationary value $1/2 \log(2)$ at large times, even though the finite-size oscillations are still strong at $\ell=200$. 

Our results are compatible with the experimental observations of Ref.~\cite{karch2025probing}, where DQPTs are detected at early times using the subsystem LE after a quench from the N\'eel state, whereas for large times the subsystem LE reaches a stationary value. Since the initial state considered in~\cite{karch2025probing} is a translation-invariant product state, both SLE and RLE are identical, and our results apply. 

\subsection{Sub-hydrodynamic regime and dynamical quantum phase transitions for the dimer and XY quenches}

For the dimer and the XY quenches introduced in Secs.~\ref{sec:dimer-quench} and~\ref{sec:xy-ground}, it is much more involved to derive analytical results in the sub-hydrodynamic regime, because the structure of the oscillating terms  is extremely intricate. In the next section, we study both quenches in the hydrodynamic regime, where these oscillations can be neglected. Here, we numerically probe the early-time dynamics in both quenches to study potential DQPTs, or lack thereof. 

An important quantity in the quench dynamics is the mode occupation number $n_k$, which is the average number of quasiparticles with momenta $k$ produced during the quench. In our quenches of interest, we have 
\begin{equation}\label{eq:nk}
(2n_k-1) =
\begin{cases}
  0, & \text{N\'eel}, \\
  \cos (k), & \text{Dimer}, \\
  \cos \Delta_k, & \text{XY},
\end{cases}
\end{equation}
where $\cos \Delta_k$ is given in Eq.~\eqref{eq:quenchparameters}. 

In general, we conjecture that the logarithmic RLE in the sub-hydrodynamic regime contains a term $\int \dd k \mu_k$ with
\begin{equation}
    \mu_k = \log(2n_k +\eE^{2\ir t \epsilon_k}),
\end{equation}
 where $\epsilon_k=\cos(k)$ for the N\'eel and dimer quenches, whereas it is given in Eq.~\eqref{eq:quenchparameters} for the XY quench. For instance, for the N\'eel quench we have $\mu_k  =\log(1+\eE^{2\ir t \cos(k)}) $, which is exactly the term discussed in Eq.~\eqref{eq:mukNeel}. 
 
\begin{figure}
    \centering
    \includegraphics[width=0.7\linewidth]{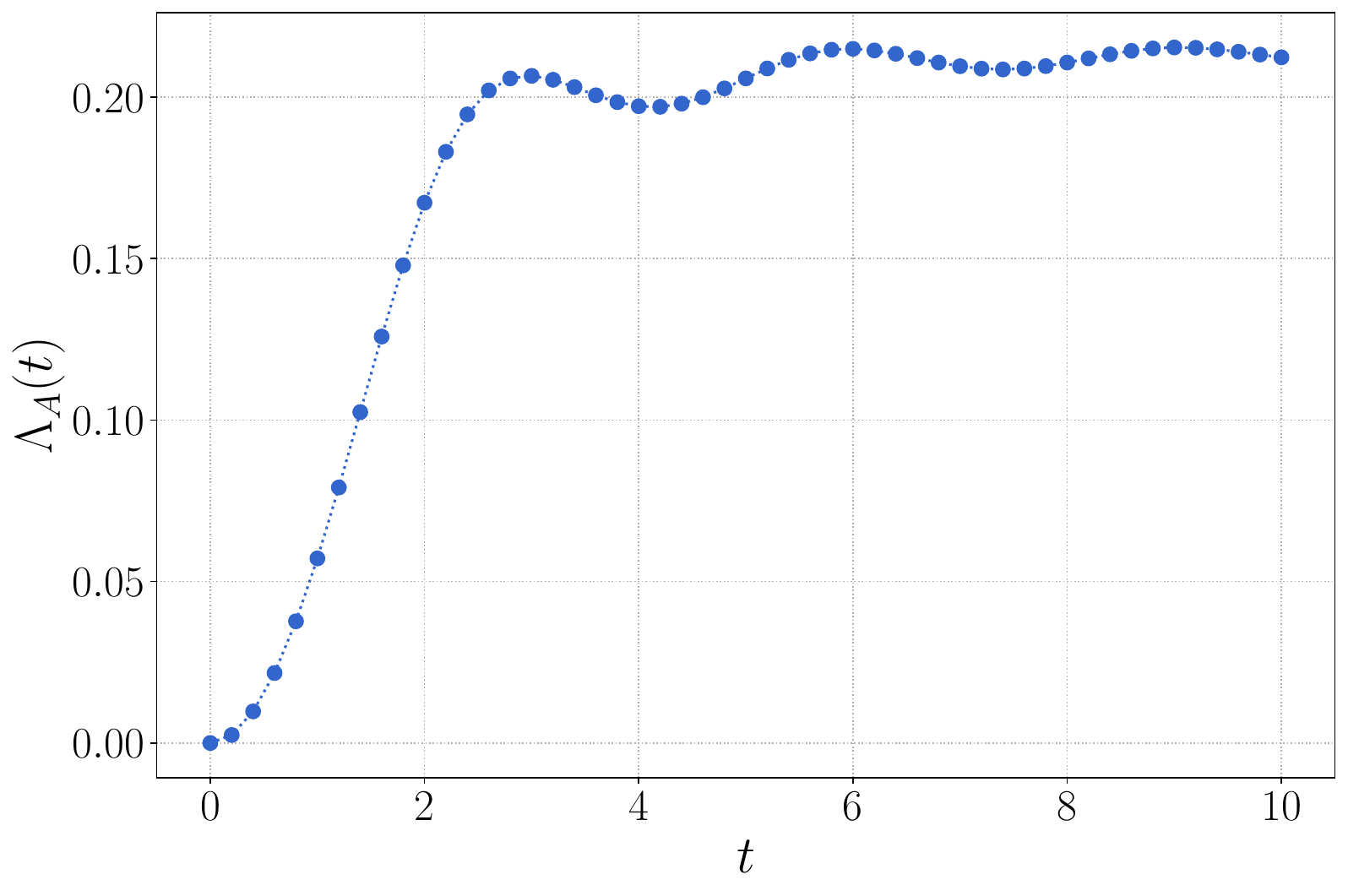}
    \caption{Logarithmic RLE in the XX chain from the dimer state with $\ell=200$ as a function of $t$. The symbols are obtained by exact numerical diagonalization.}
    \label{fig:LRLEDimEarly}
\end{figure}

 We observe that $\mu_k$ can produce non-analytic behaviors for times 
 \begin{equation}
     t = (2m+1)\frac{\pi}{2\epsilon_{k^*}}, \qquad n_{k^*} = \frac 12,
 \end{equation}
where the momentum $k^*$ is defined such that $n_{k^*} = 1/2$, and $m$ is an integer. We thus find that DQPT can occur at times $t = (m+1/2)t^*$ with $t^* = \pi/\epsilon_{k^*}$. This is consistent with the result $t^*=\pi$ discussed above for the N\'eel quench (with the additional condition that the non-analytic contribution form $k^*$ is not compensated by other terms in the integral, which gives $k^*=0$ and $\epsilon_{k^*}=1$). Interestingly, for the dimer quench we have $t^* = \infty$, implying the absence of DQPT for this quench protocol. 
In Fig.~\ref{fig:LRLEDimEarly} we display  exact numerical results for the logarithmic RLE for the dimer quench state at early times. Since $t^*=\infty$, we plot the logarithmic RLE as a function of $t$. We observe some finite-size oscillations, but these are not sharp enough to correspond to non-analyticites. These numerical results thus confirm the absence of DQPT for the dimer quench. Notice however, as we discuss in the next section, that 
in the hydrodynamic limit $t,\ell\to\infty$ with $t/\ell$ fixed the time-derivative of the RLE exhibits a nontrivial 
singularity structure, due to the presence of an infinite number of nested lightcones contributing to the dynamics.

\begin{figure}
    \centering
     \includegraphics[width=0.47\linewidth]{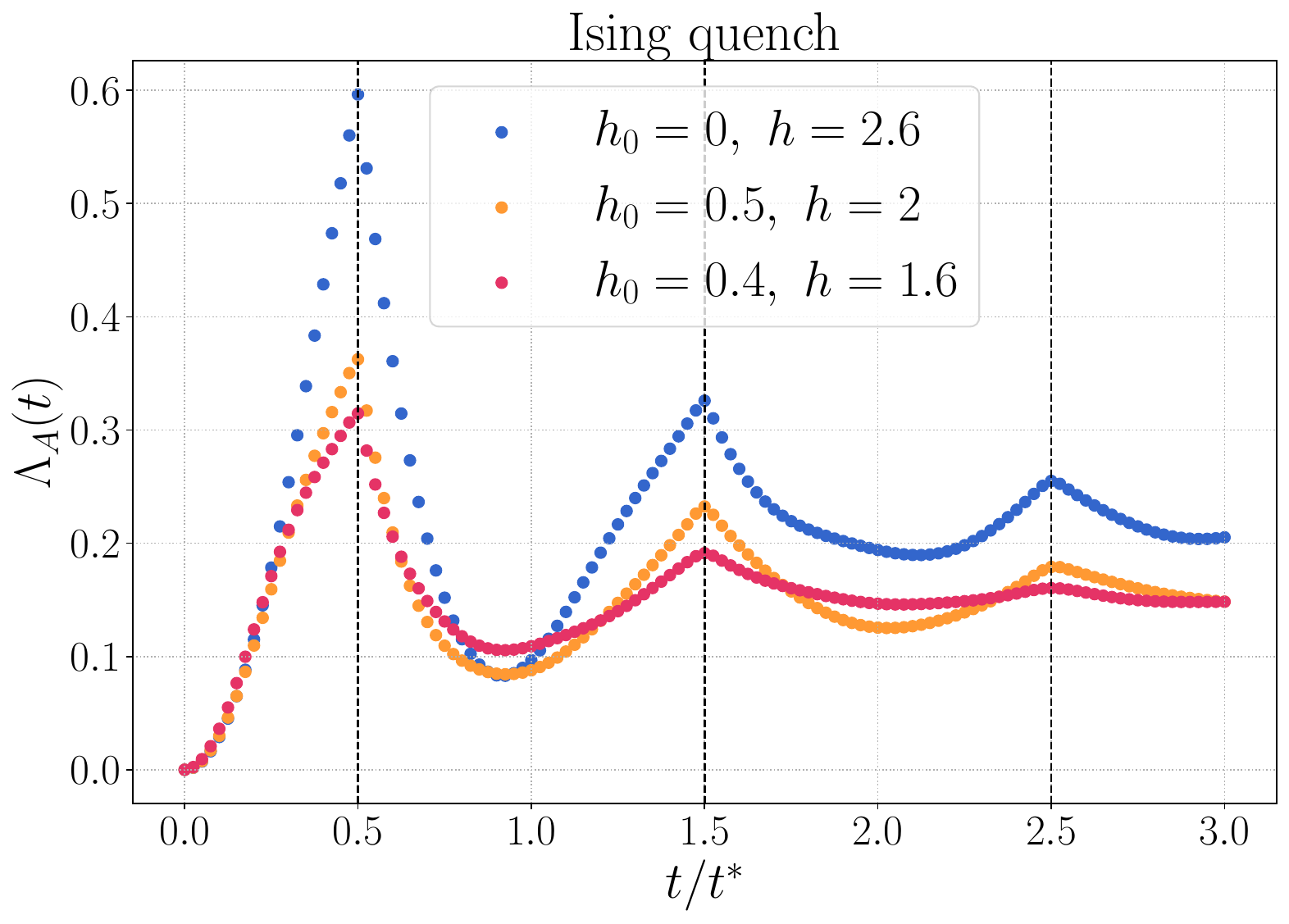}
     \includegraphics[width=0.47\linewidth]{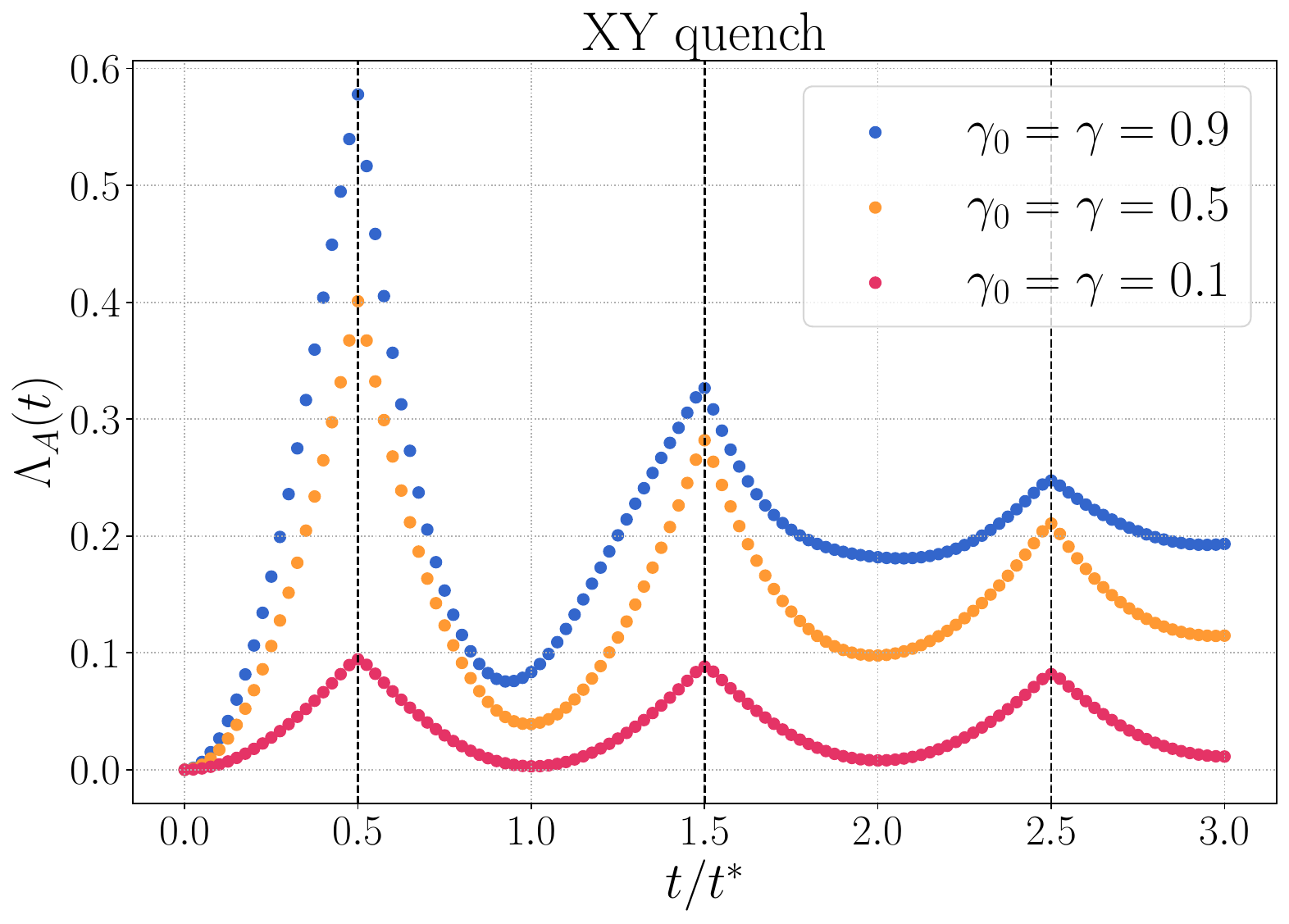}
    \caption{Logarithmic RLE in the Ising model, i.e., $\gamma_0=\gamma=1$ (left) and in the XY model with $h_0=0$ and $h=2.6$ (right) as a function of $t/t^*$ with $t^*$ obtained from the main text, for various values of the quench parameters and $\ell=200$. The symbols are obtained by exact numerical diagonalization. We observe that DQPTs occur at $t=(m+1/2)t^*$.}
    \label{fig:RLEXYDQPT}
\end{figure}

For the XY model, our conjecture suggests that we have $t^* = \pi/\epsilon_{k^*}$ with 
\begin{equation}\label{eq:ekstar}
    \epsilon_{k^*} = \sqrt{(h-\cos (k^*))^2+\gamma^2 -\gamma^2\cos^2(k^*)}
\end{equation}
and $ \cos (k^*)$ is such that $n_{k^*}=1/2$, or $\cos\Delta_{k^*}=0$. With Eq.~\eqref{eq:quenchparameters}, we thus have
\begin{equation}\label{eq:coskstar}
    \cos (k^*) = \frac{h+h_0-\sqrt{(h+h_0)^2-4(1-\gamma \gamma_0)(hh_0+\gamma \gamma_0)}}{2(1-\gamma \gamma_0)}
\end{equation}
where we restrict ourselves to $h,h_0\geqslant 0$ without loss of generality. 
In particular, for the quantum Ising model, corresponding to $\gamma=\gamma_0 = 1$, we recover the critical time $t^*=\pi/\sqrt{\big(h-\frac{1+h_0 h}{h_0+h}\big)^2+1-\big(\frac{1+h_0 h}{h_0+h}\big)^2}$ obtained for the full LE in Ref.~\cite{heyl2013dynamical}. In Fig.~\ref{fig:RLEXYDQPT}, we display the early-time logarithmic RLE for various quenches in the XY model, and observe non-analyticities at $t=(m+1/2)t^*$, where $t^*$ is obtained from Eqs.~\eqref{eq:ekstar} and \eqref{eq:coskstar}, as expected. 

\subsection{Hydrodynamic regime for the dimer and the XY quenches}
\label{sec:dimer-hydro}

Having discussed the sub-hydrodynamic regime, which might host dynamical singularities, we now derive the QPP for the dimer and the XY quenches. Precisely, in the following we derive the QPP for the dimer quench, and then we conjecture the result for the XY quench. 
As for the N\'eel quench, the key ingredients are the moments $\mathcal{M}_n=\mathrm{Tr}(J_A(0)J_A(t))^n$. 
Let us start with considering the Fourier transform of the fermionic correlation matrix $C_{nm}$ for the dimer quench,
\begin{equation}
\label{eq:C-corr}
C_{n,m}=\langle c^\dagger_n c_m\rangle=
\int_{-\pi}^\pi \frac{\dd k}{2\pi}\eE^{-ik(m-n)}
[\hat D_k(t)]_{a,b}, \qquad a,b=1,2,
\end{equation}
where we defined 
\begin{equation}
\hat D_k(t)=
\frac{1}{2}\left(\begin{array}{cc}
1-\sin(k)\sin(2t\cos(k)) &  -\cos(k)-\ir\sin(k)\cos(2t\cos(k)) \\
 -\cos(k)+\ir\sin(k)\cos(2t\cos(k)) & 1+\sin(k)\sin(2t\cos(k)) \\
\end{array}
\right).
\end{equation}
The different entries of the matrix $\hat D_k$ correspond to different 
parity of the indices $n,m$. Precisely, odd $n,m$ and even $n,m$ correspond to $[\hat D_k]_{1,1}$ and $[\hat D_k]_{2,2}$, respectively. One has 
$[\hat D_k]_{1,2}$ and $[\hat D_k]_{2,1}$  for $n$ odd and $m$ even and 
for $n$ even and $m$ odd, respectively. One can check that the correlation matrix given in Eq.~\eqref{eq:C-corr} is identical to Eq.~\eqref{eq:JAdim}.
The Fourier transform of the correlator $J_{n,m}=2C_{n,m}-\delta_{n,m}$ is $\hat J_k=2\hat D_k-\mathbb{I}$.  In the following we are interested 
in computing terms of the form $\mathrm{Tr}(\hat J_{k_1}(0)\hat J_{k_1}(t)\hat J_{k_2}(0)\hat J_{k_2}(t)\cdots \hat J_{k_n}(0)\hat J_{k_n}(t))$. To this goal, here we employ $\hat J_k(t)$ defined as 
\begin{equation}
\label{eq:J-f-dimer}
\hat J_k(t)=\left(\begin{array}{cc}
\cos(k) & \ir \eE^{2\ir t\cos(k)}\sin(k)\\
-\ir \eE^{-2\ir t\cos(k)}\sin(k) & -\cos(k)
\end{array}\right).
\end{equation}
One can check that although Eq.~\eqref{eq:J-f-dimer} is slightly different from the $\hat J_k$ obtained from Eq.~\eqref{eq:C-corr}, it gives the same results in the hydrodynamic limit. Notice also that Eq.~\eqref{eq:J-f-dimer} is obtained from the correlation matrix between the modes $c_k$ and $c_{k-\pi}$, with $k\in[-\pi,0]$, 
which corresponds to the quasiparticles forming the entangled pairs responsible for the growth of the entanglement entropies. 
Actually, let us consider a generalization of Eq.~\eqref{eq:J-f-dimer} as  
\begin{equation}
\label{eq:J-f-dimer-1}
\hat J_k(t)=\left(\begin{array}{cc}
f(k+\pi) & \eE^{2\ir t\epsilon_k}g(k+\pi)\\
\eE^{-2\ir t\epsilon_k}g(k) & f(k)
\end{array}\right),
\end{equation}
where $f(k)$ and $g(k)$ are functions. 
The dimer quench corresponds to the choice $f(k)=-\cos(k)$ 
and $g(k)=-\ir\sin(k)$. The fact that the global state 
is pure implies that 
the eigenvalues of $\hat J_k$ are $\pm1$. 
This implies in turn that the functions must satisfy $f(k)f(k+\pi)-g(k)g(k+\pi)=-1$. 

Let consider the moments $\mathcal{M}_n$. 
To do that we employ the trivial identity
\begin{equation}
\label{eq:id-1}
\sum_{m=1}^\ell \eE^{\ir m k}=\frac{\ell}{4}\int_{-1}^1 \dd\xi 
w(k) \eE^{\ir(\ell\xi+\ell+1)k/2},\quad w(k)=\frac{k}{\sin(k/2)},
\end{equation}
which follows directly from Eq.~\eqref{eq:sumexp}.
In order to simplify the calculations, in the following we relabel the sites of the chain, defining a two-site unit cell. Indeed, in the thermodynamic limit odd-even effects should be negligible.  This means that the indices in~\eqref{eq:C-corr} become cell indices. The cell indices are defined in the interval $[1,\ell/2]$. To lighten the notation we perform the calculations with $\ell$, and rescale all the lengths by a factor $1/2$ in the final formulas for $\mathcal{M}_n$. This implies that $\ell\to\ell/2$ and $v_k\to v_k/2$. 
With these conventions and Eq.~\eqref{eq:id-1}, we rewrite $\mathcal{M}_n$ as 
\begin{equation}
\label{eq:Mn}
\mathcal{M}_n=\left(\frac{\ell}{4}\right)^n
\prod_{j=1}^n \left(\int_{-\pi}^{\pi}\frac{\dd k_j}{2\pi} \int_{-1}^1\dd\xi_j \right)
\eE^{\ir\ell\sum_{p=1}^n\xi_p(k_{p-1}-k_p)/2}\prod_{j=1}^n w(k_{j-1}-k_j)\mathrm{Tr}\left(\prod_{j=1}^n \hat J_{k_j}(0)\hat J_{k_j}(t)\right).
\end{equation}
Here the trace in the last term is the trace of a two-by-two matrix.  
We can change variables to the $\zeta_j$ defined as 
\begin{equation}
    \begin{cases}
  \zeta_0=\xi_1 &\\ 
  \zeta_j=\xi_{j+1}-\xi_j,& j\in[1,n-1].
\end{cases}
\end{equation}
After noticing that the integrand in Eq.~\eqref{eq:Mn} 
does not depend on $\zeta_0$, we obtain 
\begin{equation}
\label{eq:Mn-1}
\mathcal{M}_n=\left(\frac{\ell}{4}\right)^n
\prod_{j=1}^n \left(\int_{-\pi}^{\pi}\frac{\dd k_j}{2\pi} \int_{-1}^1\dd\zeta_j\right)
\eE^{\ir\ell\sum_{p=1}^{n-1}\zeta_p(k_{p}-k_n)/2}
\mu(\{\zeta_j\})\prod_{j=1}^n w(k_{j-1}-k_j)\mathrm{Tr}\left(\prod_{j=1}^n \hat J_{k_j}(0)\hat J_{k_j}(t)\right),
\end{equation}
where $\mu$ results from the integration over 
$\zeta_0$, and it is given as 
\begin{equation}
\label{eq:mu-def}
\mu(\{\zeta_j\})=\max\left[0,\min_{j\in[0,n]}\left(1+\sum_{k=1}^j\zeta_k\right)+\min_{j\in[0,n]}\left(1-\sum_{k=1}^j\zeta_k\right)\right].
\end{equation}
The idea is now that in the scaling limit $\ell,t\to\infty$ with the ratio $t/\ell$ fixed one 
can use the stationary phase approximation. The strategy is to treat the $2n-2$ dimensional integral over $k_1,k_2,\dots, k_{n-1}$  and $\zeta_1,\zeta_2,\dots, \zeta_{n-1}$ by using the stationary phase. 
 The stationary phase approximation states that~\cite{wong2001asymptotic} 
\begin{equation}
\label{eq:statio}
\int_{\mathcal{D}} \dd^N x \ p(\vec x)\eE^{\ir\ell q(\vec x)}
\rightarrow\Big(\frac{2\pi}{\ell}\Big)^{N/2}p(\vec
x_0)|\mathrm{det}\textrm{Hess}(q(\vec x_0))|^{-1/2}\exp\Big[\ir\ell q(\vec
x_0)+\ir\pi\frac{\sigma_A}{4}\Big], 
\end{equation}
where $\mathcal{D}$ is the integration domain, $N$ is the number of variable, and $p(\vec{x}),q(\vec{x})$ are well-behaved functions. In the right-hand side $\vec{x}_0$ 
is the stationary point satisfying $\partial_{x_j}q(\vec{x})=0$, $\textrm{Hess}(q(\vec x_0))$ is the Hessian of $q(\vec{x})$ evaluated at the stationary point, and $\sigma_A$ its signature, i.e., the difference in the number of positive and negative eigenvalues of the Hessian matrix. 

First, in 
the hydrodynamic limit all the quasimomenta $k_j$ become equal, as it 
is clear from the stationarity condition with respect to the variables $\zeta_j$ in the phase factor in Eq.~\eqref{eq:Mn-1}. The stationarity condition implies that $w(k_{j-1}-k_j)\to 2$. Moreover, we can replace 
$k_j\to k$ in the integrand except in the phase factors, which are needed to impose stationarity with respect to $k_j$. To proceed, we can expand the trace in Eq.~\eqref{eq:Mn-1}. We anticipate that the stationary behavior of $\mathcal{M}_n$ will depend in a nontrivial way on $n$. To illustrate that, let us first consider the case with $n=2$. By expanding the trace one obtains different types of phase factors. One can verify  that only the terms with 
vanishing phase in the limit in which the momenta $k_j$ become equal contribute. 
For $n=2$, by keeping only the terms that survive in the hydrodynamic limit, we obtain  
\begin{multline}
\label{eq:n2}
\mathrm{Tr}(J_{k_1}(0)J_{k_1}(t)J_{k_2}(0)J_{k_2}(t))=
f^4(k)+f^4(k+\pi)+2f(k)f(k+\pi)g(k)g(k+\pi)\\
+(\eE^{2 \ir t\epsilon_{k_1}-2 \ir t\epsilon_{k_2}}+\eE^{2 \ir t\epsilon_{k_2}-2 \ir t\epsilon_{k_1}})f(k)f(k+\pi)g(k)g(k+\pi)
\end{multline}
where the first row contains the terms with no phase factors, whereas the second one contains the only terms 
with a vanishing phase in the limit $k_1\to k_2$. 
Now, when performing the stationary phase approximation on the first term we obtain that $\zeta_j=0$ for any $j$, 
which implies that $\mu=2$ (see Eq.~\eqref{eq:mu-def}). One can apply Eq.~\eqref{eq:statio} to the integrals in $\zeta_1,\zeta_2,\dots,\zeta_n$, and $k_1,k_2,\dots,k_{n-1}$ in $\M_n$. For the case with $n=2$ the term with no phase factors in Eq.~\eqref{eq:n2} gives the trivial stationary point $k_j=k_n$ for any $j$ and $\zeta_j=0$. This gives $\mu=2$. Thus, Eq.~\eqref{eq:statio} applied to the integrals in $k_1,\zeta_1$, after integrating over $k_2$, gives 
\begin{equation}
\label{eq:vol}
\ell\left(f^4(k)+f^4(k+\pi)+2f(k)f(k+\pi)g(k)g(k+\pi)
\right).
\end{equation}
The two terms with the phase factors in Eq.~\eqref{eq:n2} 
have the nontrivial stationary points 
\begin{equation}
\label{eq:statio-2}
k_1=k_2,\quad \zeta_1=\pm 4\epsilon_k't/\ell,
\end{equation}
where the $\pm$ is for the two different phases in Eq.~\eqref{eq:n2}. Both stationary points give 
the same result because $\mu$ 
is invariant under sign change of $\zeta_j$. 
One has $\mu=2\max(1-2|v_k|t/\ell,0)$, where we introduced the group velocity $v_k = \epsilon_k'$. 
The Hessian gives $|\det \textrm{Hess}|=2^{-2n+2}$,
and one obtains the contribution
\begin{equation}
\label{eq:kin0}
2\max(\ell-2|v_k|t,0) f(k)f(k+\pi)g(k)g(k+\pi).
\end{equation}
Notice that all the phase factors disappear at the stationary point, as expected. Let us observe that for $n=2$ there are two possible terms, i.e., the volume-law term~\eqref{eq:vol} and the time-dependent term~\eqref{eq:kin0}, which contains information on the propagating quasiparticles because it depends on their group velocities~$v_k$. Notice that~\eqref{eq:kin0} is the number of entangled pairs of quasiparticles that at time $t$ are in $A$. One can use the trivial identity 
\begin{equation}
\max(\ell-2|v_k|t,0)=\ell-\min(2|v_k|t,\ell),
\end{equation}
which allows us to rewrite~\eqref{eq:kin0} in terms of $\min(2|v_k|t,\ell)$, which is the number of pairs of entangled quasiparticles shared between the subsystem $A$ and the rest. 

To proceed for general $n$, we observe that since the time-dependent phase factors arising from the trace in Eq.~\eqref{eq:Mn} have to disappear at the stationary point, in the expansion of $\mathrm{Tr}(\prod_j \hat J_{k_j}(0)\hat J_{k_j}(t))$ only the phase factors with an equal number of $\epsilon_k$ 
and $-\epsilon_k$ have to be considered. Hence, for $n=3$, only phase factors of the form $\eE^{\ir(\epsilon_{k_j}t-\epsilon_{k_{j'}}t)}$ have to be considered. This means that for $n=3$ 
the dynamics of the quasiparticles is similar to~\eqref{eq:kin0}. From the expansion of $\mathrm{Tr}(\prod_j \hat J_{k_j}(0)\hat J_{k_j}(t))$ we obtain the quasiparticles contribution as 
\begin{equation}
3(f^2(k+\pi)+f^2(k))g(k)g(k+\pi)(f(k)f(k+\pi)+g(k)g(k+\pi)). 
\end{equation}
The volume-law contribution to $\mathcal{M}_3$ is obtained from the  terms 
without phase factors. Actually, we verified that  the  volume-terms can be easily obtained for generic $n$ as 
\begin{equation}
\mathrm{Tr} {M}^n,\quad\mathrm{with} \,\,{M}=\left(\begin{array}{cc}
f^2(k+\pi) & f(k+\pi)g(k+\pi)\\
f(k)g(k) & f^2(k)
\end{array}\right). 
\end{equation}
Moreover, for generic $n$ the terms with phase factors of the form 
$\eE^{\ir(\epsilon_{k_j}t-\epsilon_{k_{j'}}t)}$ are obtained from the 
coefficient of the term $xy$ in the expansion of 
\begin{equation}
\mathrm{Tr}\widetilde{M}^n,\quad\widetilde{M}=\left(\begin{array}{cc}
f^2(k+\pi)+x g(k)g(k+\pi) & x f(x)g(k+\pi)+ f(k+\pi)g(k+\pi)\\
f(k)g(k)+y f(k+\pi)g(k) & f^2(k)+g(k)g(k+\pi)
\end{array}\right).
\end{equation}

However, we now show that for $n>4$ contributions different from~\eqref{eq:kin0} 
can appear, depending on the precise functional form of $f(k)$ and $g(k)$ in Eq.~\eqref{eq:J-f-dimer-1}. 
For instance, let us consider the terms of the form $\eE^{\ir\epsilon_{k_j}t+\ir\epsilon_{k_{j'}}t-\ir\epsilon_{k_{j''}}t-
\ir\epsilon_{k_{j'''}}t}$. By requiring stationarity with respect to $k_j$, one obtains the conditions 
\begin{equation}
\label{eq:sys}
\partial_{k_j}\ell\left[\sum_{l=1}^{n-1}\zeta_l\frac{k_{l}-k_n}{2}+2\ir t\sum_{j}{\sigma_j} \epsilon_{k_{j}}\right]=0,
\end{equation}
where $\sigma_j\in \mathcal{P}(\{+,+,-,-\})$, and $\mathcal{P}(\{+,+,-,-\})$ denotes the permutation of the signs $\{+,+,-,-\}$ of the energies $\epsilon_{k_{j}}$ appearing in the phase factor. 
Let us consider the alternating permutation  $+,-,+,-$. 
The system of equations~\eqref{eq:sys} gives 
\begin{equation}
\zeta_j=\pm(-1)^j4\epsilon_k't/\ell,
\end{equation}
which gives the kinematic factor $2\max(1-2|v_k|t/\ell,0)$. The other alternating permutation $-,+,-,+$ gives the same result, and the quasiparticle contribution is 
\begin{equation}
f^2(k)f^2(k+\pi)g^2(k)g^2(k+\pi).
\end{equation}
Let us now consider the remaining four permutations of $\{+,+,-,-\}$, which all have at least a pair of equal signs with consecutive labels. Let us 
focus on $\{+,+,-,-\}$. Now, it is straightforward to check that the system~\eqref{eq:sys} has 
the solution
\begin{equation}
\zeta_1=\zeta_2=-\zeta_3=4\epsilon_k't/\ell,
\end{equation}
which gives now 
\begin{equation}
\label{eq:kin}
\mu=2\max(1-4|v_k|t/\ell,0).
\end{equation}
The kinematic factor in Eq.~\eqref{eq:kin} is unusual, as it would correspond to pairs of entangled quasiparticles 
having twice the group velocity $v_k$. We checked that all the four sign permutations 
give the same result. The contributions of the quasiparticles to $\mathcal{M}_4$ is  
\begin{equation}
\label{eq:four}
f(k)f(k+\pi)g^3(k)g^3(k+\pi). 
\end{equation}
Crucially, by inspecting the solutions of the system~\eqref{eq:sys}, one can check new 
types of kinematic terms appear upon increasing $n$. Precisely, the moment $\mathcal{M}_n$ is 
expected to give rise to all the kinematic terms $\max(1-2j|v_k|t/\ell,0)$, 
with $j\in [1,\lfloor n/2\rfloor]$. To determine which kinematic factor appears, one has to 
count the number of consecutive $+$ or $-$ in the string of signs in $\mathcal{P}(\{+,+,\dots,-,-\})$. 
We checked that the permutations with $j$ consecutive equal signs give rise to the 
kinematic factor $\max(1-2j|v_k|t/\ell,0)$. Although it is possible to determine 
the hydrodynamic formula for $\mathcal{M}_n$ for arbitrary $n$ and for arbitrary $\hat J_k(t)$ given in Eq.~\eqref{eq:J-f-dimer-1}, 
the results become  cumbersome for large $n$. However, as it is clear from~ Eq.\eqref{eq:four}, if 
$f(k)=0$ the unusual the contribution of the ``unusual'' entangled pairs vanishes. This is the 
case for the N\'eel quench, for which only the usual entangled pairs, with contributions of the form given in~\eqref{eq:kin0}, contribute, see Eq.~\eqref{eq:LRLENeelHydro}.

We now restrict ourselves to the case of the dimer  quench and 
for $n\leqslant 7$. The final result for  $\mathcal{M}_2$ reads  
\begin{subequations}\label{eq:MnHydroDimer}
\begin{equation}
\label{eq:m2_dimer}
\mathcal{M}_2=\int\frac{\dd k}{2\pi}\Big\{\max\left(\ell-2|v_k| t,0\right)(-\cos^2(k)\sin^2(k))+\ell\cos^2(k)\cos(2k)\Big\},
\end{equation}
A tedious although straightforward calculation gives 
\begin{equation}
\label{eq:m3_dimer}
\mathcal{M}_3=\int\frac{\dd k}{2\pi}\Big\{\max\left(\ell-2|v_k |t,0\right)\frac 38(- \sin(2k)\sin(4k))+\ell \cos^3(k)\cos(3k)\Big\},
\end{equation}
\begin{multline}
\label{eq:m4_dimer}
\mathcal{M}_4=\int\frac{\dd k}{2\pi}\Big\{\max(\ell-2|v_k|t,0)
\frac{1}{8}\cos^2(k)\sin^2(k)(-7-16\cos(2k)-25\cos(4k))\\-
2\max(\ell-4|v_k|t,0)\cos^2(k)\sin^6(k)
+\ell\cos^4(k)\cos(4k)\Big\},
\end{multline}
\begin{multline}
\label{eq:m5_dimer}
\mathcal{M}_5=\int\frac{\dd k}{2\pi}\Big\{\max(\ell-2|v_k|t,0)\frac{5}{4}\cos^4(k)\sin^2(k)(-5+6\cos(2k)-9\cos(4k))
\\-5\max(\ell-4|v_k|t,0)\cos^2(k)\sin^6(k)(1+2\cos(2k))+\ell\cos^5(k)\cos(5k)\Big\},
\end{multline}
\begin{multline}
\label{eq:m6_dimer}
\mathcal{M}_6=\int\frac{\dd k}{2\pi}\Big\{\max(\ell-2|v_k|t,0)\frac{1}{32}\cos^4(k)\sin^2(k)(-2-179\cos(2k)+2\cos(4k)-301\cos(6k))
\\-\frac{3}{4}\max(\ell-4|v_k|t,0)\cos^2(k)\sin^6(k)(23+40\cos(2k)+21\cos(4k))
\\-3\max(\ell-6|v_k|t,0)\cos^2(k)\sin^{10}(k)+\ell\cos(6k)\cos^6(k)\Big\},
\end{multline}
\begin{multline}
\label{eq:m7_dimer}
\mathcal{M}_7=\int\frac{\dd k}{2\pi}\Big\{\max(\ell-2|v_k|t,0)\frac{7}{64}\cos^4(k)\sin^2(k)(-19-22\cos(2k)-40\cos(4k)-42\cos(6k)-69\cos(8k))
\\-\frac{7}{2}\max(\ell-4|v_k|t,0)\cos^4(k)\sin^6(k)(17+16\cos(2k)+23\cos(4k))
\\-7\max(\ell-6|v_k|t,0)\cos^2(k)\sin^{10}(k)(2+3\cos(2k))+\ell\cos(7k)\cos^7(k)\Big\}.
\end{multline}
\end{subequations}

\begin{figure}[t]
\centering
\includegraphics[width=.7\linewidth]{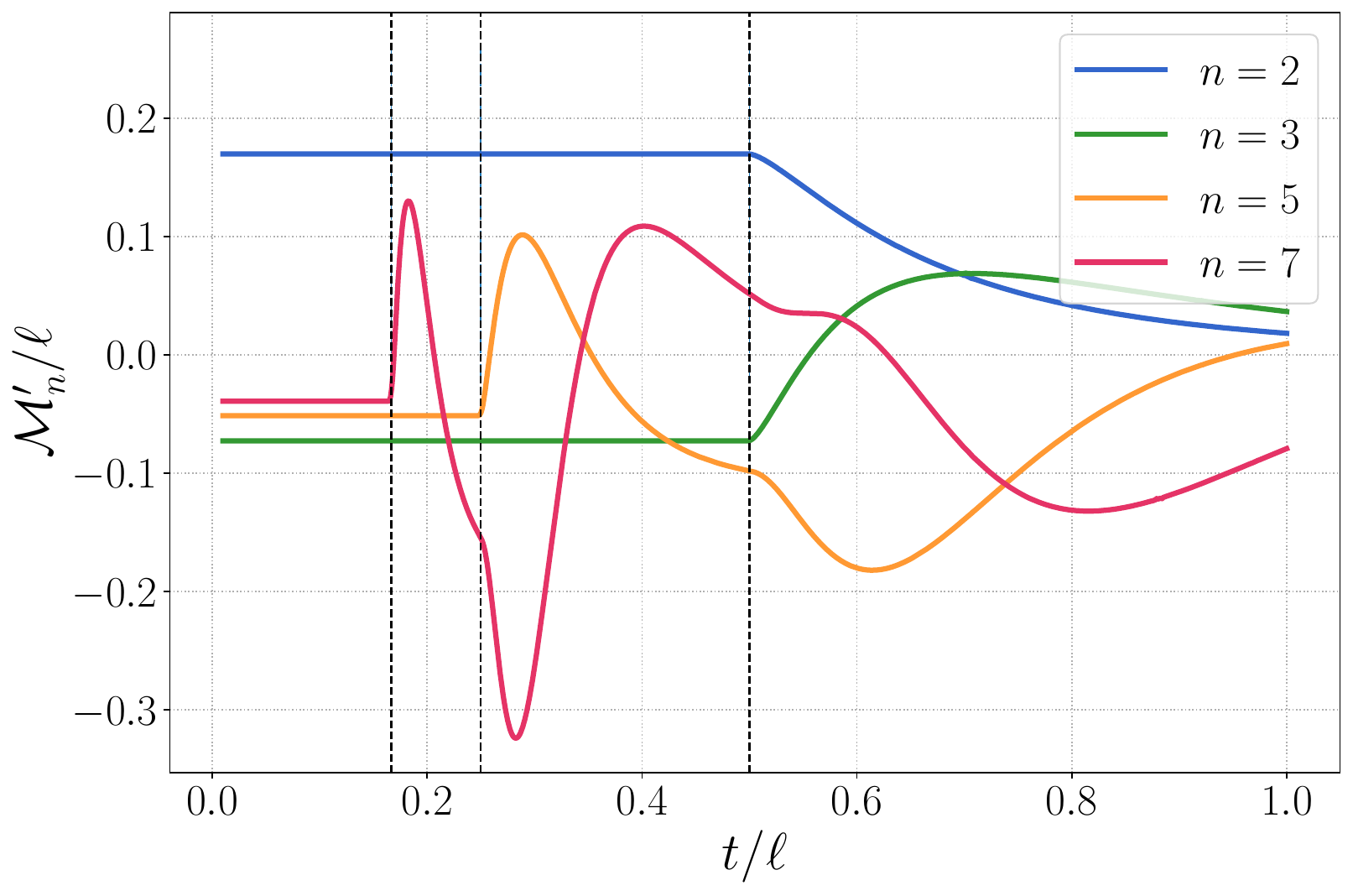}
\caption{Derivative $M'_n = \dd M_n/\dd t$ of the moments as a function of $t/\ell$ for the quench from the dimer state. The solid lines are the analytical results of Eq.~\eqref{eq:MnHydroDimer}, and the dotted vertical lines are located at $t/\ell = 1/(2j)$ with $j=1,2,3$. }
\label{fig:dimermn}
\end{figure}

As anticipated, $\mathcal{M}_n$ contains pairs of quasiparticles with maximum velocity up to $\lfloor n/2 \rfloor \max_k(|v_k|)$. For the dimer quench, we have $v_k =\sin(k)$, and hence $\max_k(|v_k|)\equiv v_{\rm max} =1$.
This implies that the derivative with respect to time $\mathcal{M}_n'$ exhibits 
cusp-like singularities at $t/\ell=1/(2j)$. We confirm that in Fig.~\ref{fig:dimermn} 
plotting $\mathcal{M}'_n/\ell$ for $n\in[2,7]$ as a function of $t/\ell$. 
The vertical lines are the values $t/\ell=1/(2j v_{\rm max})$, with $j=1,2,3$. 
To obtain the total RLE, we sum over an infinite number of $\M_n$. Hence, we have an infinite structure of nested lightcones with seemingly arbitrary large velocities. This unusual structure however does not contradict the Lieb-Robinson bound \cite{lieb1972finite}, because we consider products of operators with full support in the subsystem $A$. A similar feature arises for the full-counting statistics after a quench \cite{groha2018full}. 

Finally, let us discuss the quasiparticle picture for the dynamics of the RLE for the XY quench. 
An ab initio derivation of the QPP for the XY quench is quite cumbersome. However, we observe that 
the quasiparticle density $n_k$ (see Eq.~\eqref{eq:nk}) for the XY quench is the same as that for the 
dimer quench if one replaces $\cos(k)\to\cos\Delta_k$, where $\Delta_k$ is the Bogoliubov angle 
describing the quench. This suggests that the QPP for the dynamics of $\mathcal{M}_n = (-1)^n \Tr(\Gamma_A(0)\Gamma_A(t))^n$ after the XY quench is obtained from Eq.~\eqref{eq:MnHydroDimer} after replacing 
$\cos(k)\to\cos\Delta_k$ and multiplying the integrals by an overall factor 2, because the dimension of the correlation matrices is $2\ell\times 2\ell$, instead of $\ell \times \ell$ for the dimer quench.

\subsection{Numerical benchmarks}

\begin{figure}[t]
\centering
\includegraphics[width=.47\linewidth,height=6cm]{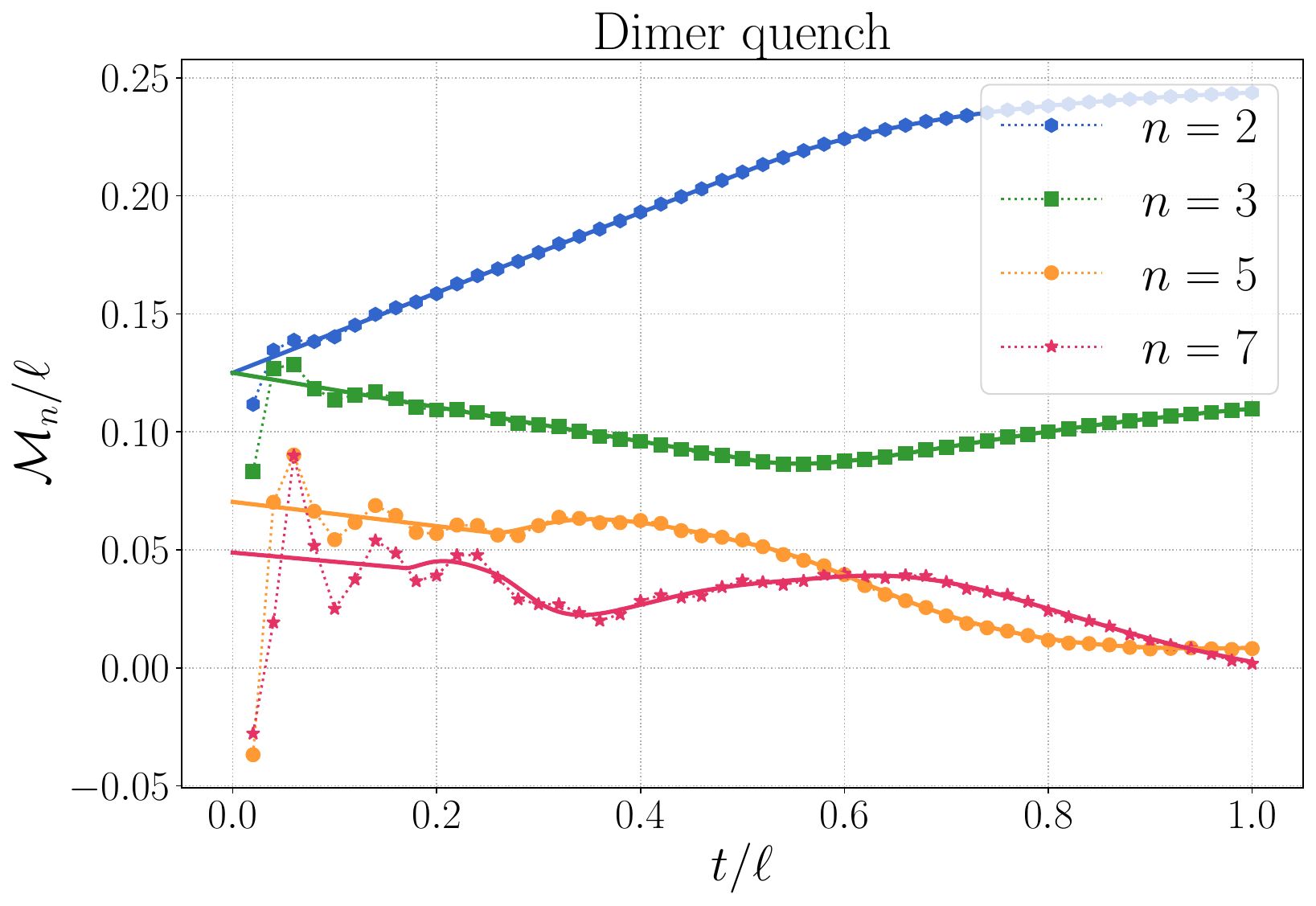}
\includegraphics[width=.47\linewidth,height=6cm]{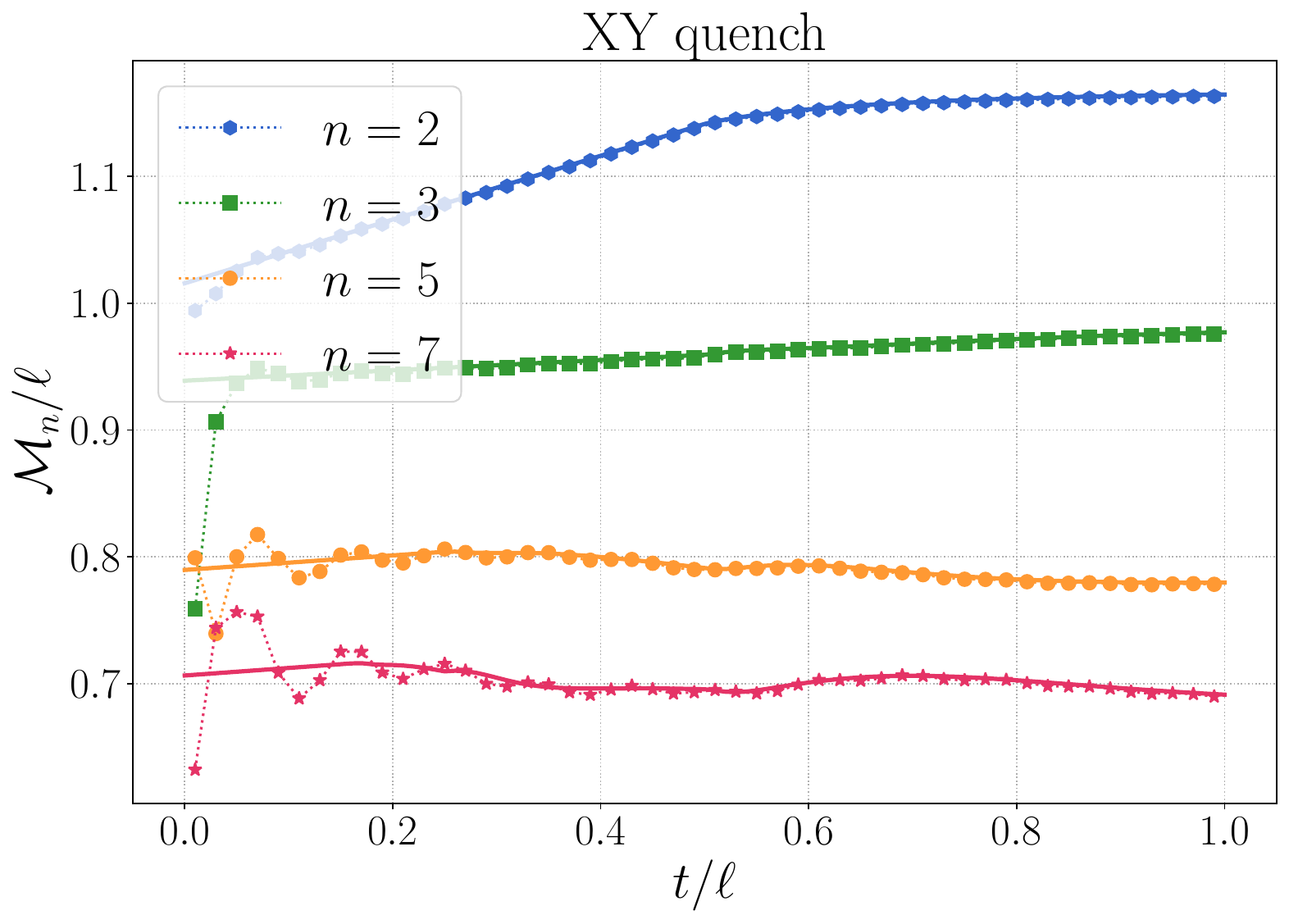}
\caption{Moments $\M_n$ for the quench from the dimer state (left) and the XY chain for $\ell=750$  (right). The initial parameters are $h_0=0.5, \ \gamma_0=0.5$ and the quenched parameters are $h=2, \ \gamma=1$. }
\label{fig:dimer_comp}
\end{figure}

Let us now discuss numerical results for the dynamics of the logarithmic RLE after the dimer quench 
and the XY quench. In Fig.~\ref{fig:dimer_comp} we show numerical results for the moments $\mathcal{M}_n$ for $n\leqslant 7$. The left and right panels are for the 
dimer and the XY quench, respectively. The full symbols are exact numerical data 
for $\mathcal{M}_n/\ell$ with $\ell=750$ and several values of $n$. The solid lines are the predictions in the hydrodynamic limit of Eq.~\eqref{eq:MnHydroDimer}. At small $t/\ell$ the data exhibit 
quite strong oscillating corrections. Still, the agreement with the theory is quite 
impressive, even for the XY quench, where the QPP remains a conjecture. 

\begin{figure}
    \centering
    \includegraphics[width=0.47\linewidth]{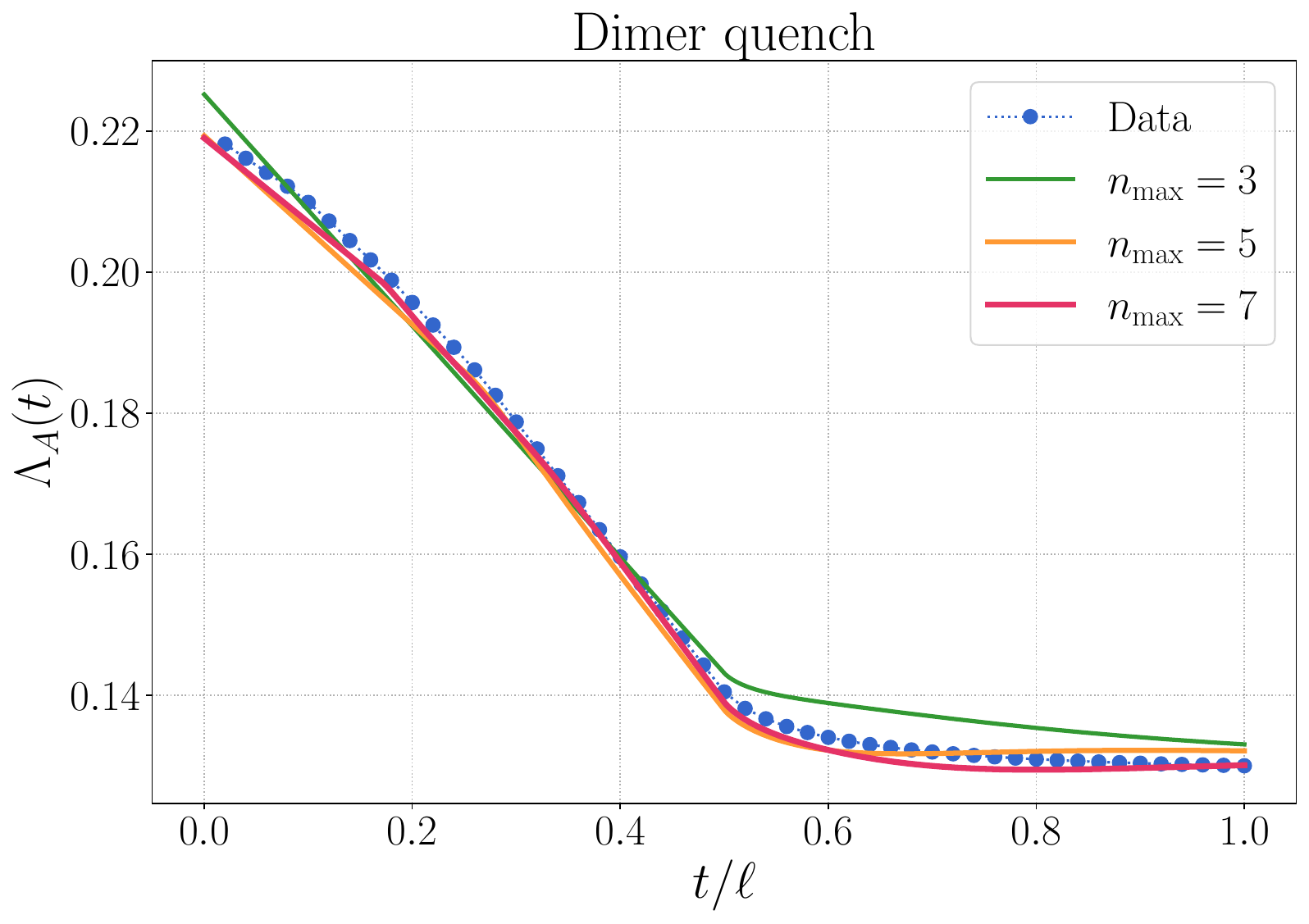}
    \includegraphics[width=0.47\linewidth]{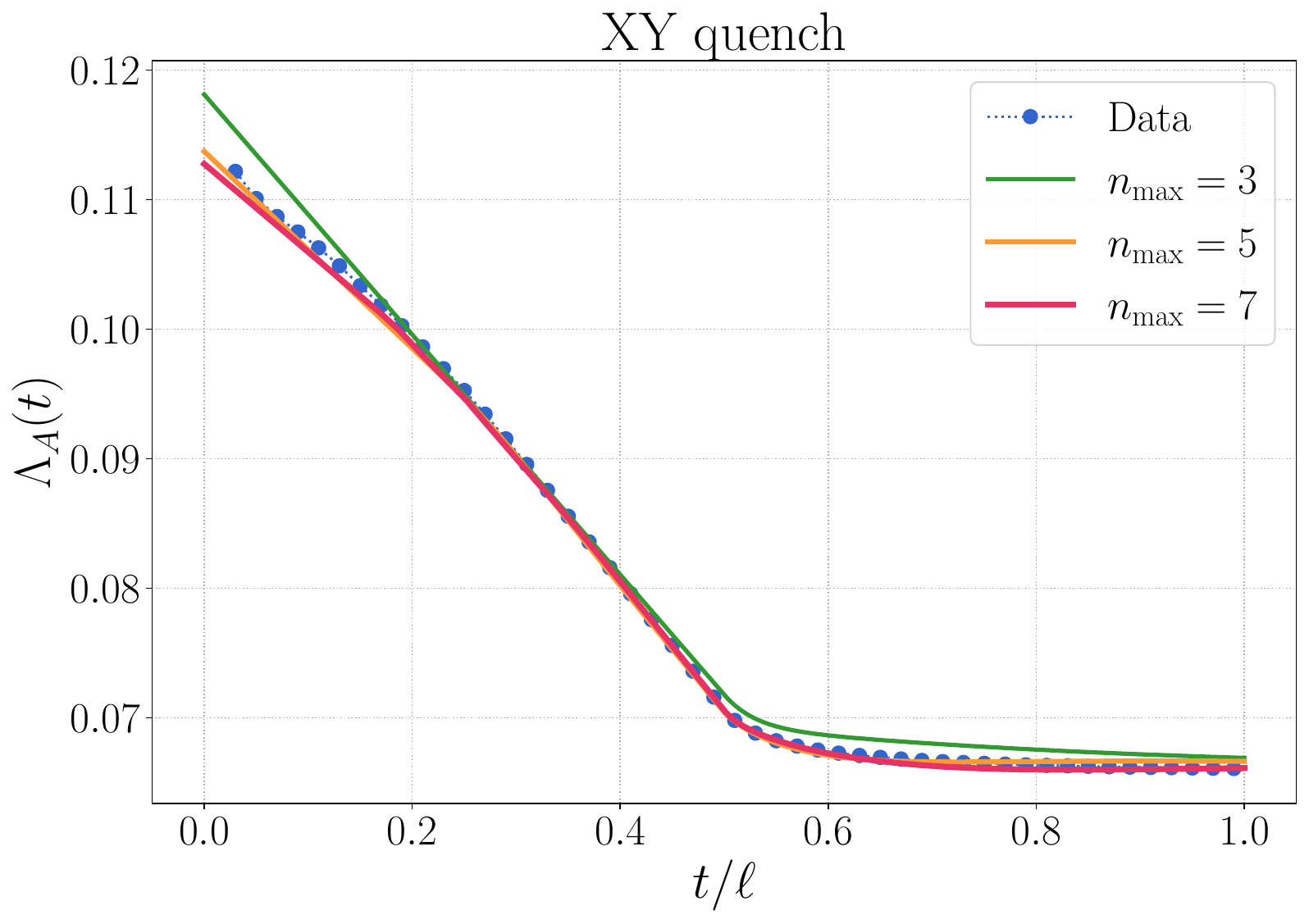}
    \caption{Logarithmic RLE for the dimer quench (left) and in the XY chain with $h_0=0.5, \ \gamma_0=0.5, \ h=2, \ \gamma=1$ (right), both with $\ell=750$ as a function of $t/\ell$. The symbols are obtained by exact diagonalization and the solid lines are the truncated theoretical prediction with various values of $n_{\rm max}$.}
    \label{fig:LRLE}
\end{figure}

\begin{figure}
    \centering
    \includegraphics[width=0.7\linewidth]{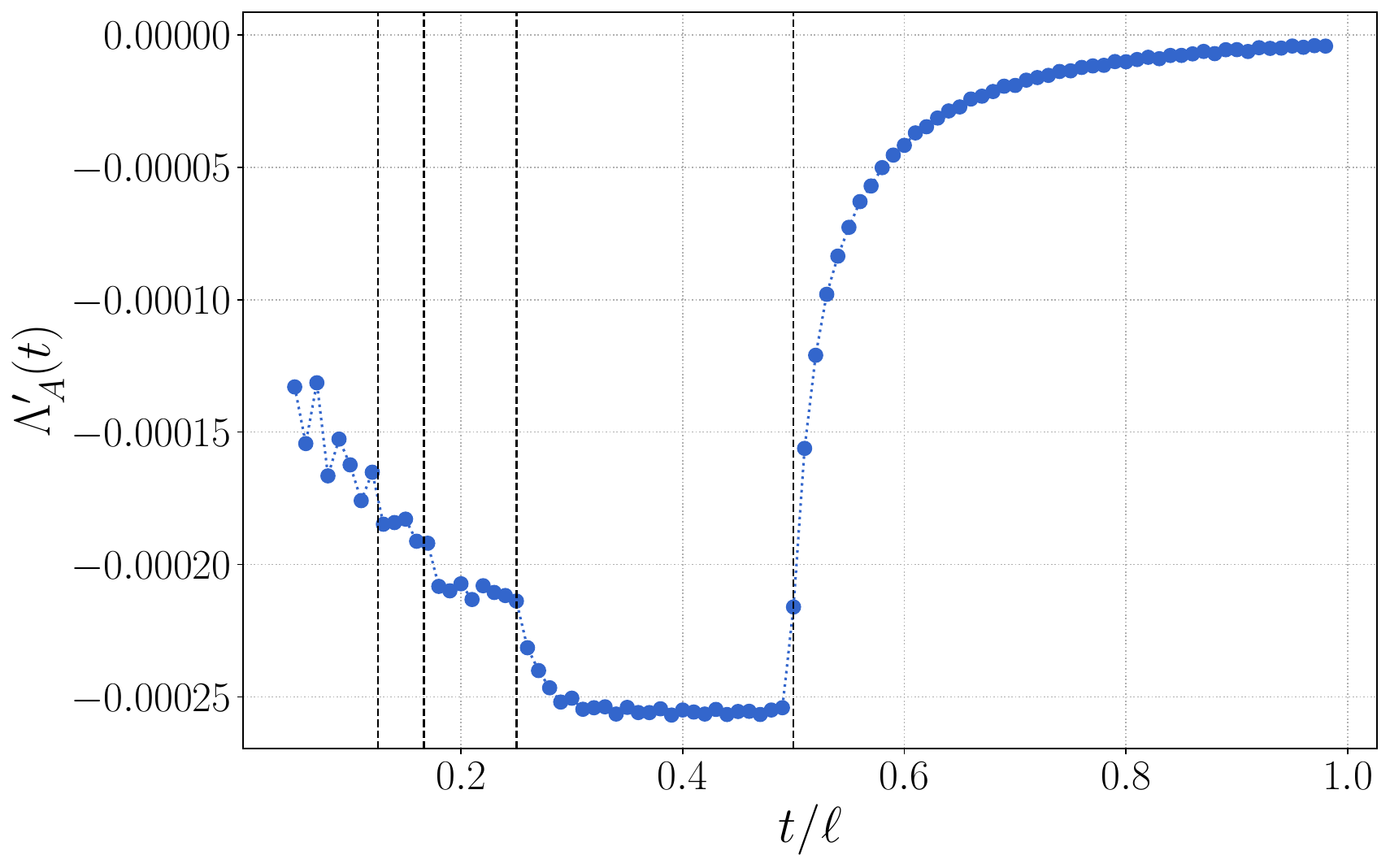}
    \caption{Time derivative of the logarithmic RLE, $\Lambda_A'(t)=\dd \Lambda_A(t)/\dd t$, for the dimer quench with $\ell=750$ as a function of $t/\ell$. The symbols are obtained by exact diagonalization and the dotted lines are located at $t/\ell=1/(2j)$ with $j=1,2,3,4$.}
    \label{fig:LRLEDT}
\end{figure}

Having the QPP for the all the moments $\mathcal{M}_n$, one can re-sum the series in Eq.~\eqref{eq:series}, and use the known QPP results for the entanglement dynamics in those quenches \cite{fc-08}, to obtain the full dynamics of the RLE. In practice, in Eq.~\eqref{eq:MnHydroDimer} we observe that $\M_n$ has a contribution of the form $\int \dd k \cos(nk)\cos(k)^n $ for $n \leqslant 7$, and we conjecture that this term is present for all $n$. The re-summation of this contribution in Eq.~\eqref{eq:series} is trivial. For the other contributions, since we do not have a generic re-summable formula for any $n$, we simply use a finite number $n_{\rm max}$ of explicit terms in the series.
In Fig.~\ref{fig:LRLE} 
we plot the logarithmic RLE as a function of $t/\ell$ for both the dimer and 
the XY quenches. The symbols are numerical data for $\ell=750$. The different lines are the results in the hydrodynamic limit obtained 
by truncating Eq.~\eqref{eq:series} up to the first $n_{\rm max}$ 
terms. As it is clear from the figure, upon increasing $n_{\rm max}$ the agreement with the numerics improves. Finally, it is interesting to investigate the effects of the nested lightcone structure on the dynamics of $\Lambda_A(t)$. To do that, in Fig.~\ref{fig:LRLEDT}  we consider the derivative $\Lambda_A'(t)$ with respect to time for the dimer quench. The symbols are numerical data for $\Lambda_A'(t)$ plotted versus $t/\ell$. The vertical dotted lines are located at $t/\ell=1/(2j)$ for $j=1,2,3,4$. The numerical data exhibit the expected staircase singularity structure.

\section{Final-state fidelity and the quantum Mpemba effect}\label{sec:FSF}

In this section we investigate the logarithmic final-state fidelity defined in Eq.~\eqref{eq:fsf} in the N\'eel, dimer and XY quenches. In particular, we show that the quasiparticle picture applies, in contrast with the case of RLE. Furthermore, we show that the FSF provides a natural tool to identify the Mpemba effect. 

\subsection{Quasiparticle picture}

\begin{figure}
    \centering
    \includegraphics[width=0.7\linewidth]{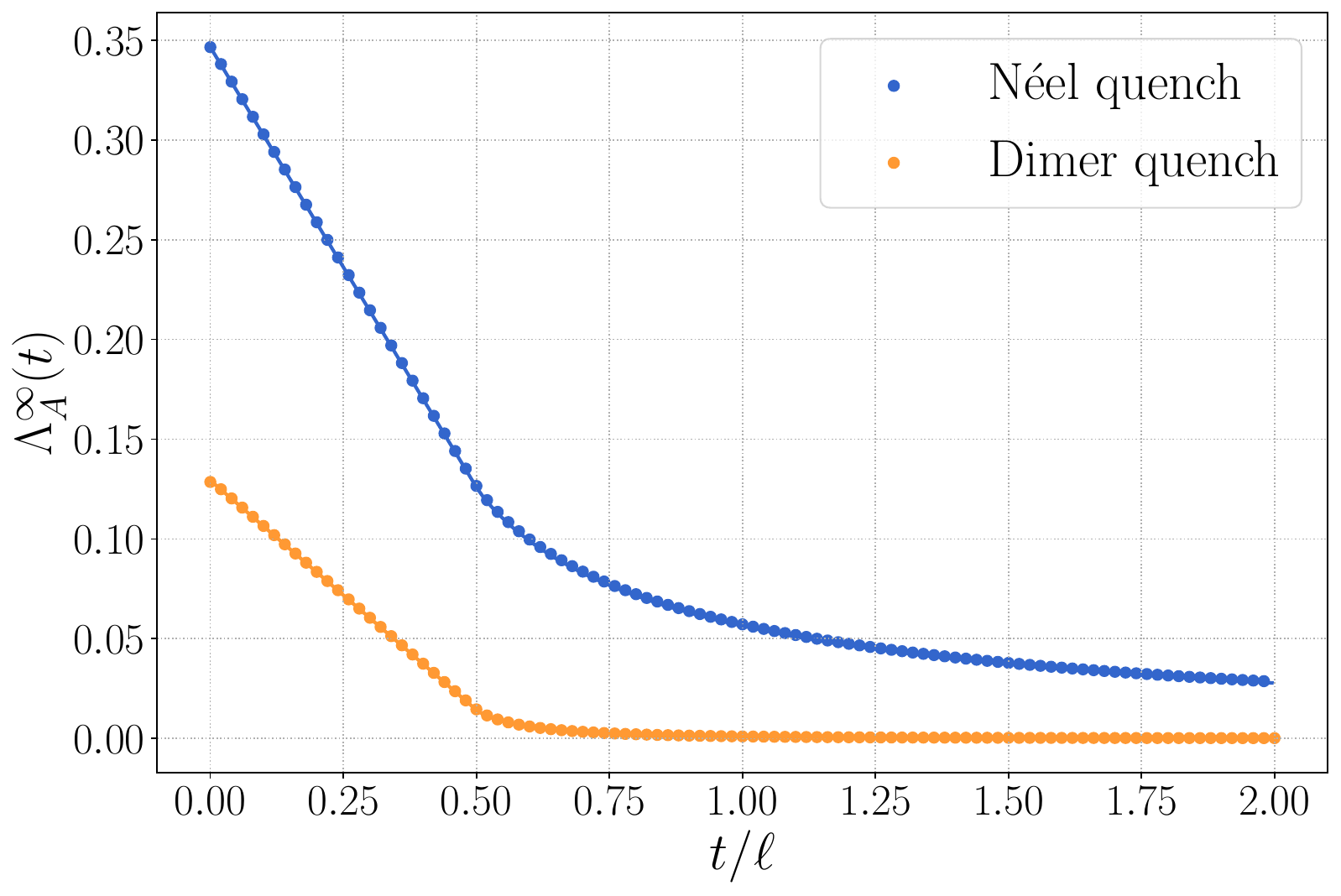}
    \caption{Logarithmic FSF for the N\'eel and the dimer quenches with $\ell=200$. The symbols are obtained by exact diagonalization and the solid lines are the theoretical predictions of Eqs.~\eqref{eq:FSFNeel} and \eqref{eq:FSFdimer}.}
    \label{fig:FSFNeeldim}
\end{figure}

In contrast with the RLE, the FSF does not exhibit singular behavior and DQPT at early times in a sub-hydrodynamic regime. We thus focus on the hydrodynamic regime where $t,\ell \to \infty$ with a fixed ratio. The analytical calculations are similar to those for the RLE presented above, the only difference being that one has to replace the initial-state correlation matrix with the stationary one.
Interestingly, stationary-state quantities make the calculations much simpler. For the N\'eel quench we find 
\begin{equation}\label{eq:FSFNeel}
\Lambda^\infty_A(t) =  \frac 12\log (2)\int_{-\pi}^{\pi} \frac{\dd k}{2\pi}  (1-\min(2|v_k| t/\ell,1) ),
\end{equation}
whereas for the dimer quench it reads
\begin{equation}\label{eq:FSFdimer}
    \Lambda^\infty_A(t) = \int_{-\pi}^\pi \frac{\dd k}{2 \pi} \left( \frac 12 \log\Big(\frac{1+\cos(k)^2}{2}\Big)-\log \Big(\frac{1+\cos(k) \eE^{\ir k}}{2}\Big) \right)\left(1-\min( 2 |v_k|t/\ell, 1)\right),
\end{equation}
with $v_k = \sin(k)$ for both quenches. We verify Eqs.~\eqref{eq:FSFNeel} and \eqref{eq:FSFdimer} in Fig.~\ref{fig:FSFNeeldim} and find a perfect agreement. 
For the XY quench we obtain 
\begin{equation}\label{eq:FSFXY}
\Lambda_A^\infty(t) = \int_{-\pi}^\pi \frac{\dd k}{2 \pi} \left( \frac 12 \log\Big(\frac{1+\cos\Delta_k^2}{2}\Big)-\log \Big(\frac{1+\cos\Delta_k \eE^{\ir \Delta_k}}{2}\Big) \right)\left(1-\min( 2 |v_k|t/\ell, 1)\right)
\end{equation}
where $v_k=\epsilon_k'$ and $\cos\Delta_k$ are given in Eq.~\eqref{eq:quenchparameters}. We verify this analytical prediction for various initial states $h_0,\gamma_0$ under the XX quench ($h=\gamma=0$) in Figs.~\ref{fig:Mpemba} and \ref{fig:Mpemba2}, and we find a perfect agreement with the numerical results. We comment more on these curves in the following section. 

Let us now discuss the general structure of the quasiparticle prediction for 
the FSF. This can be obtained within the framework of Ref.~\cite{caceffo2024fate}. To illustrate the result, let us consider the reduced density matrix $\rho_A$ of a subregion $A$ of the full system. Since we restrict ourselves to Gaussian systems, we have 
$\mathrm{Tr}G(\rho_A)=\mathrm{Tr}\mathcal{G}(J_{A})$, where $J_{A}$ is the 
real-space correlation matrix restricted to subsystem $A$, and $\mathcal{G}(x)$ is a function that implements $G(x)$ at the level of the correlation matrix. The main idea 
of Ref.~\cite{caceffo2024fate} is that for sufficiently well-behaved functions $\mathcal{G}(x)$ one can determine the hydrodynamic limit behavior of $\mathrm{Tr}G(\rho_A)$ as 
\begin{equation}
\label{eq:fabio}
\mathrm{Tr}G(\rho_A)=\sum_{x,k}\mathrm{Tr}_\mathcal{A} \mathcal{G}(\hat J_{x,k}^{(\mathcal{A})}(t)). 
\end{equation}
Here, $\hat J_{x,k}(t)$ is the correlation matrix in momentum space for the mesoscopic cell at 
position $x$. Moreover $\hat J_{x,k}(t)$ is a two-by-two matrix in the space of quasiparticles species. 
For instance, for the case of the dimer quench and the N\'eel quench there are two species $\eta^\dagger_1(k)=c^\dagger_k$ and 
$\eta^\dagger_2(k)=c^\dagger_{k-\pi}$. They have opposite group velocities and form   the entangled pairs that are responsible for the growth of the entanglement entropies. In Eq.~\eqref{eq:fabio}, $\mathcal{A}$  denotes the species in $A$. For instance, if 
only quasiparticles of species $1$ are in $A$ one has $\mathcal{A}=\{1\}$, whereas if both species are in $A$ one has 
$\mathcal{A}=\{1,2\}$. The superscript $\mathcal{A}$ in $\hat J^{(\mathcal{A})}_{x,k}(t)$ is in the space of species, and indicates that we select the rows and columns in $\hat J_{x,k}(t)$ corresponding to the quasiparticles species $\mathcal{A}$.
Since we focus on quenches from homogeneous initial states the dependence on $x$ in Eq.~\eqref{eq:fabio} disappears. 
Finally, the idea of Ref.~\cite{caceffo2024fate} is that the dependence on $t$ in $\hat J^{(\mathcal A)}_{x,k}(t)$ 
at the hydrodynamic level is due to the ballistic propagation of the quasiparticles.  Interestingly, this holds for both diagonal and off-diagonal correlations. For instance, the entry $(1,2)$ of $\hat J_{x,k}(t=0)$ encodes the correlation between the modes $1,2$ in the cell at position $x$ and time $0$. At a generic time $t$ the same entry describes the correlation between modes $1,2$ now in the cells at positions $x+v^{(1)}_kt$ and $x+v^{(2)}_kt$.  This implies that the sum over $x$ and trace over $\mathcal{A}$ are straightforward. 
For the case of two species of quasiparticles one has that $\mathcal{A}=\{1\},\{2\},\{1,2\}$, which 
correspond to the situation in which one has only quasiparticles of species $1$, $2$, or both in $A$. 
The sum over $x$ gives for the first two cases $\min(2|v_k|t,\ell)$, where we used that $v^{(1)}_k=-v^{(2)}_k$ and hence $|v^{(1)}_k|=|v^{(2)}_k| \equiv |v_k|$. 
For $\mathcal{A}=\{1,2\}$ one has $\max(\ell-2|v_k|t,0)=\ell-\min(2|v_k|t,\ell)$. The contributions $\mathrm{Tr}_\mathcal{A} \mathcal{G}(\hat J^{(\mathcal {A})}_{x,k}(t))$ are obtained by taking the initial correlator $\hat J_{x,k}(t)$ in species space and restricting to the rows and column that 
correspond to the species in $A$ at time $t$. Crucially, in doing that we replace all the phase factors with $1$. The reason is that in the hydrodynamic limit the phase factor give rise to the ballistic propagation of the quasiparticles, which is accounted for already.  To proceed, we observe that it is natural to 
extend Eq.~\eqref{eq:fabio} to functions of $\rho_A(t)\rho_A^\infty$. This suggests that 
\begin{equation}
\label{eq:fabio-1}
\mathrm{Tr}G(\rho_A(t)\rho_A^\infty)=
\sum_{x,k}\mathrm{Tr}_\mathcal{A}\mathcal{G}(\hat{\tilde J}^{(\mathcal{A})}_{x,k}(t)),
\end{equation}
where $\hat{\tilde J}_{x,k}(t)$ is constructed from 
the product $\hat J_k(t)\hat J_k(\infty)$, with 
$\hat J_k(\infty)$ obtained from $\hat J_k(t)$ by removing the time-dependent off-diagonal terms, which vanish via dephasing in the long-time limit.  Moreover, in $\hat {\tilde J}_{x,k}$ one has to set the time-dependent phases to $1$, which is equivalent to consider $\hat{\tilde J}_{x,k}(t=0)$. The time dependence in $\hat{\tilde J}_{x,k}(t)$ is taken into account by the quasiparticle propagation. Similarly as in Eq.~\eqref{eq:fabio}, the superscript $\mathcal{A}$ in Eq.~\eqref{eq:fabio-1} denotes the restriction to rows and columns that correspond to the species in~$A$. Now, we should observe that since $\hat J_k(\infty)$ is diagonal, the contributions that correspond to only one species in $A$ cancel out with the denominator in the definition of the FSF, see Eq.~\eqref{eq:FSF}. This implies that the FSF is nonzero only if both the entangled pairs are in $A$. By employing Eq.~\eqref{eq:fabio} for the denominators and Eq.~\eqref{eq:fabio-1} for the numerator in the definition of the FSF, one recovers Eqs.~\eqref{eq:FSFNeel}, \eqref{eq:FSFdimer} and \eqref{eq:FSFXY}, after using that the FSF is real. 

We illustrate this procedure for the dimer quench, where $\hat J_k$ is given in Eq.~\eqref{eq:J-f-dimer}. For $\mathcal{A}=\{1,2\}$ we have  
\begin{equation}
    \mathrm{Tr}_\mathcal{A}\mathcal{G}(\hat{\tilde J}^{(\mathcal{A})}_{x,k}(0))=-\log\det\left(\frac{\mathbb{I}+\hat J_k(0)\hat J_k(\infty)}{2}\right)=-\log\left(\frac{5+3\cos(2k)}{8}\right).
\end{equation}
For the denominators in Eq.~\eqref{eq:FSF}, we use Eq.~\eqref{eq:fabio} and find
\begin{equation}
\begin{split}
    \mathrm{Tr}_\mathcal{A}\mathcal{G}(\hat{ J}^{(\mathcal{A})}_{x,k}(\infty)) &= -\log\det\left(\frac{\mathbb{I}+\hat J_k(\infty)^2}{2}\right)=-2\log\left(\frac{3+\cos(2k)}{4}\right), \\
    \mathrm{Tr}_\mathcal{A}\mathcal{G}(\hat{ J}^{(\mathcal{A})}_{x,k}(0)) &=-\log\det\left(\frac{\mathbb{I}+\hat J_k(0)^2}{2}\right) = 0.
    \end{split}
\end{equation}
Putting everything together, we obtain the total contribution 
\begin{equation}
    \mathrm{Tr}_\mathcal{A}\mathcal{G}(\hat{\tilde J}^{(\mathcal{A})}_{x,k}(0)) -\frac 12\mathrm{Tr}_\mathcal{A}\mathcal{G}(\hat{ J}^{(\mathcal{A})}_{x,k}(\infty)) = \log \left(\frac{2(3+\cos(2k))}{5+3\cos(2k)}\right), \quad k \in [-\pi, 0]. 
\end{equation}
To integrate over the symmetric domain $k\in [-\pi, \pi]$, we simply divide by a factor 2 and get $1/2 \log(2(3+\cos(2k))/(5+3\cos(2k)))$, which coincides with the time-independent integrand in Eq.~\eqref{eq:FSFdimer}. 

Finally, we can write a formula in terms of the mode occupations $n_k$ (See Eq.~\eqref{eq:nk}) for the FSF after a generic quench in the a quadratic fermionic systems. We obtain
\begin{subequations}\label{eq:FSFQPP}
\begin{equation}
    \Lambda_A^\infty(t) = \int_{-\pi}^\pi \frac{\dd k}{2 \pi}   \left(1-\min(2|v_k| t/\ell,1)\right)\Lambda_k
\end{equation}
where $\Lambda_k$ reads
\begin{equation}
     \Lambda_k=\frac 12 \log\Big(\frac{1+(2n_k-1)^2}{2}\Big)-\log \Big(\frac{\sqrt{1+3(2n_k-1)^2}}{2}\Big) .
\end{equation}
\end{subequations}
We conjecture that this QPP formula holds in arbitrary quadratic fermionic models. The generalization to 
interacting integrable systems is an interesting open problem. It is not  clear whether a quasiparticle picture description applies in the presence of interactions.  For instance, while a quasiparticle picture can be used to understand the dynamics of the entanglement entropy~\cite{alba2017entanglement}, its generalization to the R\'enyi entropies proved to be hard~\cite{alba2017quench,alba2017renyi,bertini2022growth} (see Ref.~\cite{bertini2023non} for a similar result for the full-counting statistics). 

\subsection{Quantum Mpemba effect with symmetry restoration}

Let us consider a nonequilibrium situation where the Hamiltonian $H$ generating the time evolution possesses a certain symmetry. For simplicity, we assume a $U(1)$ symmetry with charge $Q$, namely $[H,Q]=0$. We thus expect that, irrespective of the initial state, all stationary states also possess that symmetry, and thus satisfy $[\rho_A^{\infty},Q_A]=0$, where $Q_A$ is the charge restricted to subsystem $A$. Given an initial state $|\Psi_0\rangle$ which breaks the symmetry, $[|\Psi_0\rangle \langle \Psi_0|, Q]\neq 0$, a natural question is to quantify how fast is the symmetry restored during the time evolution, or how fast does the system thermalize and reaches its stationary state. Certain situations give rise to the quantum Mpemba effect, where dynamics from initial states which strongly break the symmetry restore it faster than dynamics from initial states closer to a symmetric state \cite{ares2023entanglement,ares2025quantum}. The quantum Mpemba effect has been investigated theoretically and experimentally using various measures, such as the entanglement asymmetry\cite{ares2023entanglement,joshi2024observing,murciano2024entanglement}, the trace distance \cite{aharony2024inverse,zhang2025observation}, the Frobenius distance \cite{carollo2021exponentially} or the relative entropy \cite{moroder2024thermodynamics,ares2025simpler}. Quantum fidelities akin to the FSF have also been used to detect quantum Mpemba effects in two dimensional bosonic systems \cite{yamashika2025quenching}. For a recent review on quantum Mpemba effects, see Ref.~\cite{ares2025quantum}.  

\begin{figure}
    \centering
    \includegraphics[width=0.47\linewidth]{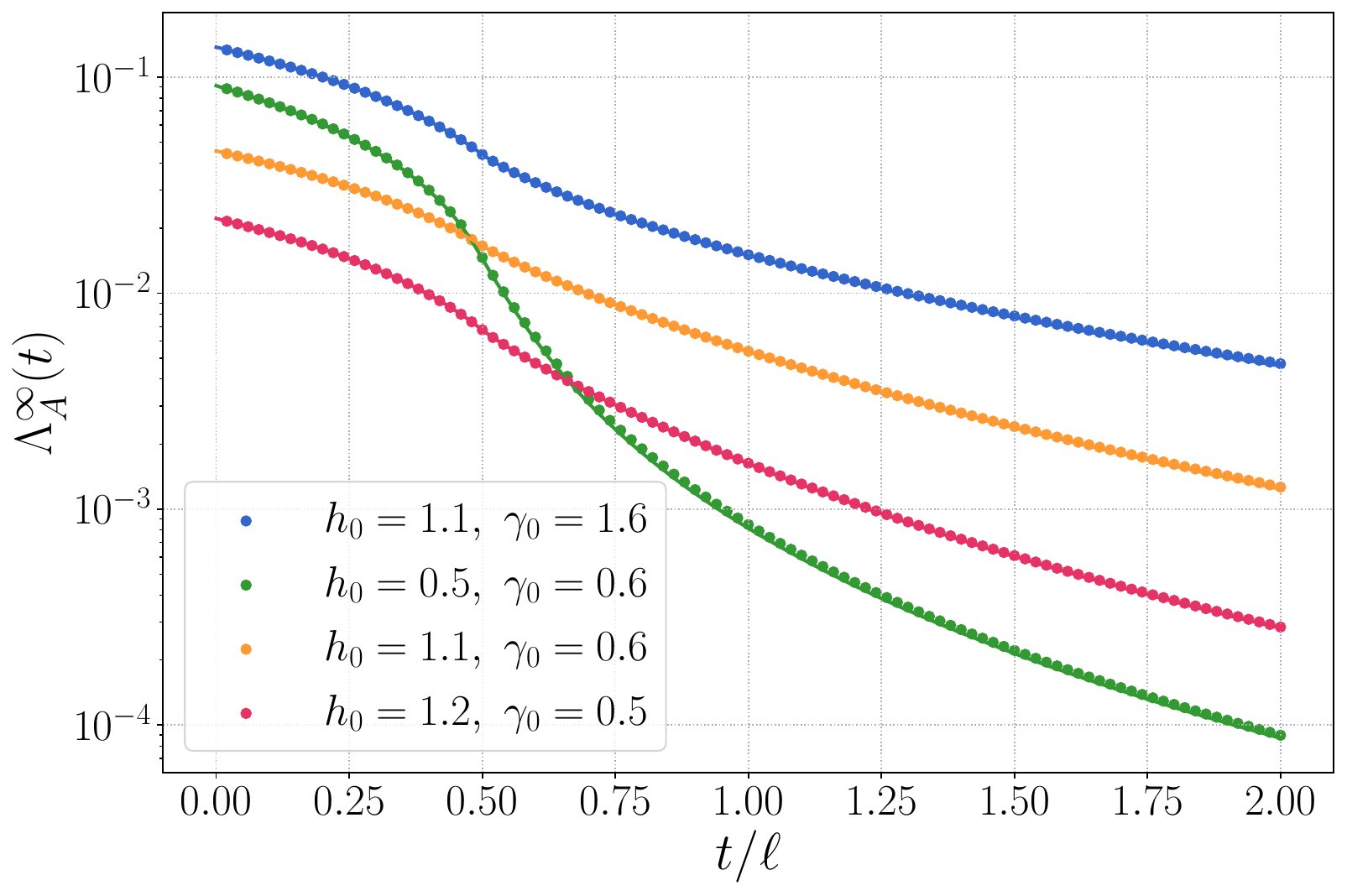}
    \includegraphics[width=0.47\linewidth]{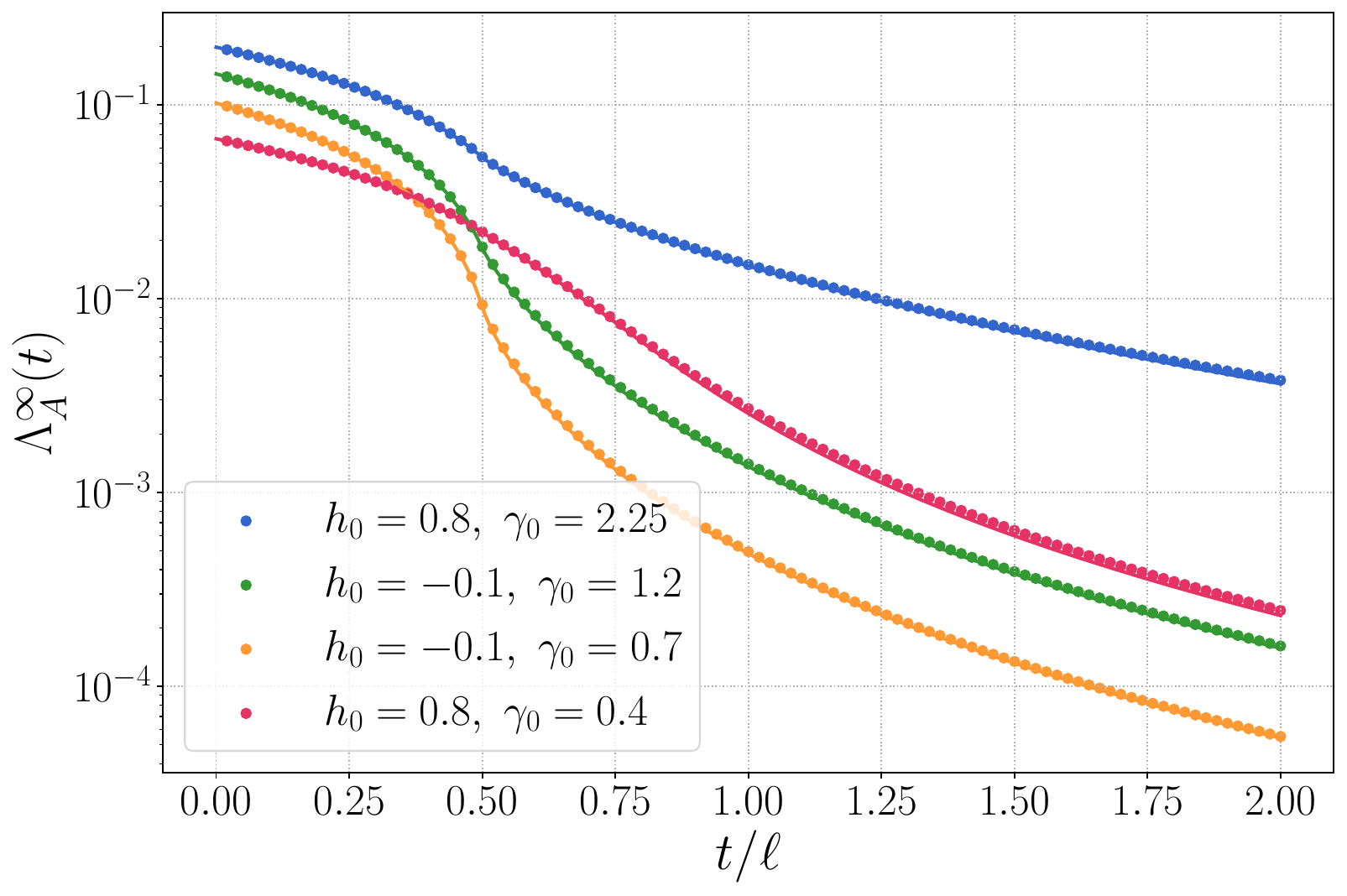}
    \caption{Logarithmic FSF in the XX chain ($h=\gamma=0)$ with $\ell=200$ in logarithmic scale. The initial states are ground states of the XY chain with various values of $h_0, \gamma_0$. The symbols are obtained by exact numerical diagonalization, and the solid lines are the analytical prediction of Eq.~\eqref{eq:FSFXY}.}
    \label{fig:Mpemba}
\end{figure}

\begin{figure}
    \centering
    \includegraphics[width=0.7\linewidth]{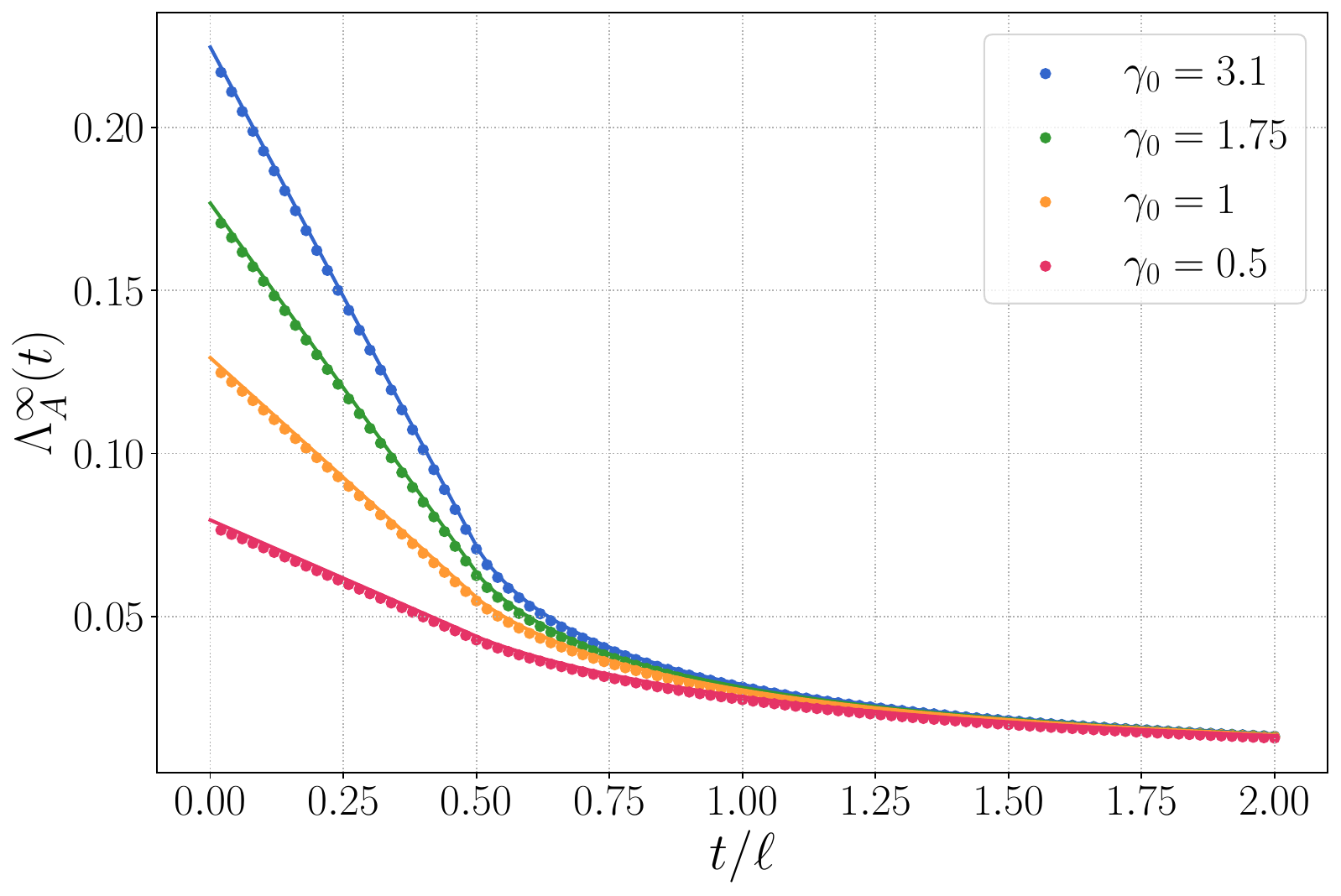}
    \caption{Logarithmic FSF in the XX chain ($h=\gamma=0)$ with $\ell=200$. The initial states are ground states of the critical XY chain with $h_0=1$ and various values of $\gamma_0$. No crossings occurs in this case. The symbols are obtained by exact numerical diagonalization, and the solid lines are the analytical prediction of Eq.~\eqref{eq:FSFXY}.}
    \label{fig:Mpemba2}
\end{figure}

By design, the FSF is a powerful tool to probe thermalization in quantum systems out of equilibrium, and in particular the rate at which the system reaches its stationary state after a quench. Moreover, reduced fidelities have a strong experimental relevance, as we already discussed for the RLE. We thus expect the logarithmic FSF to be a natural quantity for investigating quantum Mpemba effects, both analytically and experimentally. However, as opposed to the entanglement asymmetry which directly quantifies symmetry breaking and its restoration, the FSF is an indicator of the relaxation towards a symmetric stationary state. These two quantities are thus complementary in the study of thermalization in quantum systems. To test this claim, we consider quenches in the XX chain ($h=\gamma=0$) from ground states of the XY chain with various values of $h_0,\gamma_0$. The XX Hamiltonian possesses a $U(1)$ symmetry, where the associated charge is simply the fermion number, whereas ground states of the XY chain with arbitrary anisotropy $\gamma_0$ break this symmetry. Quantum Mpemba effects, or lack thereof, have been diagnosed in this quench protocol via the entanglement asymmetry \cite{murciano2024entanglement}. In Figs.~\ref{fig:Mpemba} and \ref{fig:Mpemba2}, we consider the same quenches as in Ref.~\cite{murciano2024entanglement}, and obtain exactly the same qualitative results. In particular, in Fig.~\ref{fig:Mpemba} the crossings indicate that certain states initially further away from their stationary values thermalize faster than other more symmetric initial states, hence corresponding to a quantum Mpemba effect. In contrast, in Fig.~\ref{fig:Mpemba2} we display situations where such crossings do not appear, and where more symmetric initial states do thermalize faster than less symmetric ones. 

\subsection{Quantum Mpemba effect without symmetry restoration}

Mpemba effects are not always related to a symmetry restoration. Indeed, there are quench protocols where dynamics from initial states further away from the stationary state relax faster, even though the Hamiltonian driving the time evolution does not have symmetries which are broken by the initial states. Such Mpemba effects were for instance observed in the Ising quench using the relative entropy between the time-evolved RDM and the stationary state \cite{ares2025simpler}.

\begin{figure}
    \centering
    \includegraphics[width=0.7\linewidth]{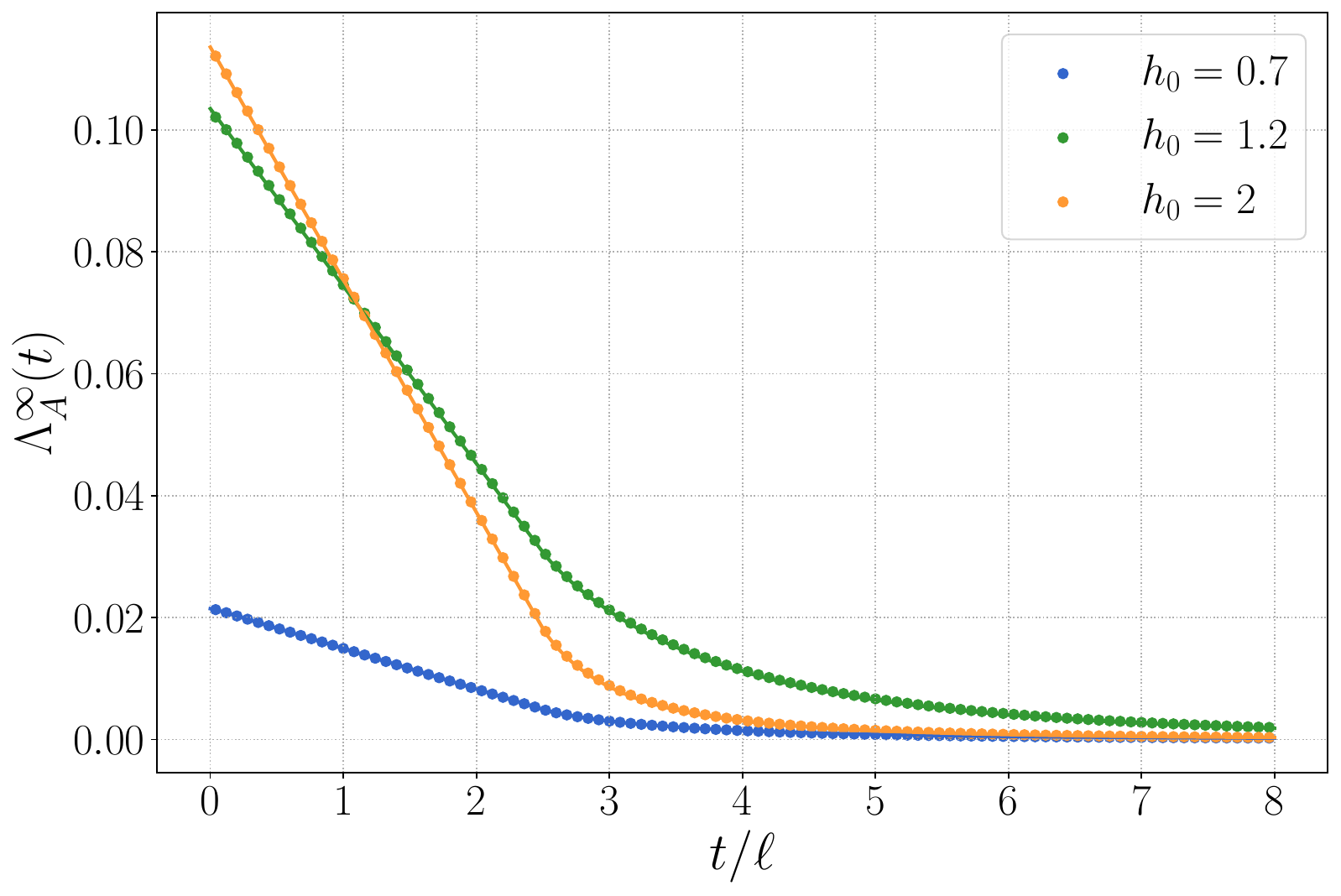}
    \caption{Logarithmic FSF in the Ising chain ($\gamma=1$) with $h=0.2$ and $\ell=200$. The initial states are ground states of the Ising chain ($\gamma_0=1$) with various values of $h_0$. The symbols are obtained by exact numerical diagonalization, and the solid lines are the analytical prediction of Eq.~\eqref{eq:FSFXY}.}
    \label{fig:Mpemba3}
\end{figure}

As already mentioned, the FSF is a probe of how fast a system relaxes after a quench. In particular, it is still well-defined and relevant even in contexts where there is no symmetry restoration in the dynamics, as opposed to the entanglement asymmetry. We thus expect it to be able to diagnose Mpemba effects, even without symmetry restoration. To test this, we consider the same quench protocols as in Ref.~\cite{ares2025simpler}: 
a quench to the Ising chain with $\gamma=1$ and $h=0.2$ from the Ising chain ($\gamma_0=1$) with various values of~$h_0$. We illustrate our results in Fig.~\ref{fig:Mpemba3}. As expected, we observe the same qualitative crossings (or lack thereof) as in Ref.~\cite{ares2025simpler}. Moreover, we again find a perfect agreement between the numerical results and the QPP formula \eqref{eq:FSFQPP} for the FSF.

\subsection{Conditions for Mpemba effects}

To understand our results quantitatively, it is important to investigate under which conditions Mpemba effects occurs.  
Let us consider two quenches $\mathcal{Q}_1$ and $\mathcal{Q}_2$ 
where $\mathcal{Q}_j = (h_0^{(j)},\gamma_0^{(j)}, h^{(j)}, \gamma^{(j)})$ encodes the data of the quench $j$, and $\mathcal{Q}_1\neq \mathcal{Q}_2$. We assume without loss of generality that initially we have $ \Lambda_A^{\infty,\mathcal{Q}_1}(0)> \Lambda_A^{\infty,\mathcal{Q}_2}(0)$. By definition, a quantum Mpemba effect occurs if there is a crossing time $t_C$ such that for all $t>t_C$ the inequality is reversed, $ \Lambda_A^{\infty,\mathcal{Q}_1}(t)< \Lambda_A^{\infty,\mathcal{Q}_2}(t)$. The QPP formula~\eqref{eq:FSFQPP} for the FSF is monotonically decreasing. Therefore, in the thermodynamic limit, 
the condition for a quantum Mpemba effect reduces to the existence of a finite time $0<t_C<\infty$ such that 
\begin{equation}\label{eq:tC}
     \Lambda_A^{\infty,\mathcal{Q}_1}(t_C)= \Lambda_A^{\infty,\mathcal{Q}_2}(t_C),
\end{equation}
and this condition is typically both necessary and sufficient.
There is no explicit formula for the crossing time $t_C$, but the exact QPP form of $\Lambda_A^{\infty}(t)$ (see Eq.~\eqref{eq:FSFQPP}) allows us to derive it precisely with Eq.~\eqref{eq:tC}. We stress that there are examples of quenches with multiple crossing \cite{chalas2024multiple}, where different times $t_C^{(i)}$ satisfy Eq.~\eqref{eq:tC}. In this case there is an Mpemba effect if the number of crossing times is odd.

Similarly as in Ref.~\cite{murciano2024entanglement}, where the authors derive the large-time QPP behavior of the entanglement asymmetry, we can derive the conditions for the presence of quantum Mpemba effects in the large-time limit with the FSF. 
Given two quenches $\mathcal{Q}_1$ and $\mathcal{Q}_2$, a quantum Mpemba effect occurs if there is a time~$t_I$ such that 
\begin{equation}\label{eq:cond1}
\left\{
\begin{aligned}
&\int_{-\pi}^\pi \frac{\dd k}{2\pi}\,\Lambda^{\mathcal{Q}_1}_k 
  > \int_{-\pi}^\pi \frac{\dd k}{2\pi}\,\Lambda^{\mathcal{Q}_2}_k ,
   \\[1em]
&\int_{-\hat k}^{\hat k} \frac{\dd k}{2\pi}\,(\Lambda^{\mathcal{Q}_1}_k+\Lambda^{\mathcal{Q}_1}_{k+\pi}) 
  < \int_{-\hat k}^{\hat k} \frac{\dd k}{2\pi}\,(\Lambda^{\mathcal{Q}_2}_k+\Lambda^{\mathcal{Q}_2}_{k+\pi}) ,
  \quad & t>t_I \,,
\end{aligned}
\right.
\end{equation}
where $\hat k=v^{-1}(\ell/(2t))$. For instance, $v^{-1}(x)=\arcsin(x)$ in the XX quench. The first condition in Eq.~\eqref{eq:cond1} does not depend on time and simply reflects the initial condition $ \Lambda_A^{\infty,\mathcal{Q}_1}(0)> \Lambda_A^{\infty,\mathcal{Q}_2}(0)$.
The second, however, involves the modes with momenta in $[-\hat k,\hat k]$, which are the pairs of quasiparticles that are in $A$ at time~$t$. Moreover, $\Lambda_k^{\mathcal{Q}_j}$ and $\Lambda_{k+\pi}^{\mathcal{Q}_j}$ are the 
contributions to the FSF due to the members of the generic entangled pair. Finally, since we are in the large-time limit, we only consider the slowest modes and hence neglect the contributions proportional to the velocity term $2|v_k|t$ in the integrals. Let us also stress that for the XX quench, the large-time assumption is $\ell/(2t)<1$, otherwise the momenta $\hat k$ is not well defined. In terms of the QPP interpretation, this means that we wait at least for the fastest moving quasiparticles initially created in the center of $A$ to leave the subsystem, after a time $t=v_{\rm max}\ell/2$, with $v_{\rm max}=1$ for the XX quench. 
Typically, we thus have $t_I>t_C$, because the Mpemba crossing occurring at time $t_C$ can occur for $t_C/\ell< v_{\rm max}/2$, and hence before the validity of the large-time criterion given in Eq.~\eqref{eq:cond1}. For instance, (i) in the left panel of Fig.~\ref{fig:Mpemba} for $\mathcal{Q}_1=(0.5,0.6)$ (green curve) and $\mathcal{Q}_2=(1.1,0.6)$ (orange curve) we have $t_C/\ell = 0.48$, (ii) in the right panel of Fig.~\ref{fig:Mpemba} for $\mathcal{Q}_1=(-0.1,0.7)$ (orange curve) and $\mathcal{Q}_2=(0.8,0.4)$ (pink curve) we have $t_C/\ell = 0.37$, and (iii) in the same panel for $\mathcal{Q}_1=(-0.1,1.2)$ (green curve) and $\mathcal{Q}_2=(0.8,0.4)$ (pink curve) we have $t_C/\ell = 0.48$.

\section{Conclusion}\label{sec:ccl}

In this paper, we investigated the dynamics of quantum fidelities out of equilibrium in one-dimensional free-fermion systems. Specifically, we studied reduced fidelities between the time-dependent state of a subsystem $A$ and its initial value, or the stationary state at large times. The former case corresponds to the reduced Loschmidt echo, whereas the latter defines the final-state fidelity. We focused on quench protocols where the initial states are homogeneous and Gaussian, and the dynamics is generated by the spin-$1/2$ XY Hamiltonian. Using standard algebra of Gaussian fermionic states, we expressed the reduced fidelities in terms of time-dependent correlation functions, which are known analytically in the quench protocols of interest. This allowed us to derive exact ab-initio results and to perform precise numerical simulations for the fidelities dynamics. We first considered the RLE in the sub-hydrodynamic regime, where both the time $t$ and the subsystem size $\ell$ are large, but with $t \ll \ell$. For certain quenches, we find that the RLE becomes singular periodically at times $t=(m+1/2)t^*$, signaling the presence of a DQPT. Importantly, these singularities are subleading and vanish in the large-time limit, and the critical time $t^*$ coincides with the critical time of the full-system Loschmidt echo. Moreover our conclusions are compatible with recent experimental results for the subsystem Loschmidt echo \cite{karch2025probing}, a quantity closely related to the RLE. Second, we considered the hydrodynamic regime, where both $t$ and $\ell$ are large, but with a fixed finite ratio $t/\ell$. In this limit the singular oscillating terms vanish, and we obtain a QPP-like behavior with an infinite sequence of nested lightcones. However, this surprising behavior does not contradict the Lieb-Robinson bound. Finally, we focused on the FSF, for which we find a standard QPP behavior with a single lightcone and no subleading singularities. We argue that the FSF is a natural tool to probe the thermalization of isolated quantum systems and detect quantum Mpemba effects. In contrast with other probes of the Mpemba effect, here we derived the complete time dependence of the FSF, and hence obtained precise quantitative conditions both for the presence of Mpemba effects, and for the crossing times at which they occur.

Our work paves the way for numerous exciting research directions. First, it would be interesting to consider the dynamics of RLE for dissipative systems, and in particular understand how dissipation affects the presence of DQPT and the structure of nested lightcones in the hydrodynamic regime. Another promising idea would be to consider more general initial states, possibly with spatial inhomogeneities or multiplets of quasiparticles excitations, and to generalize our QPP formulas to account for those more realistic scenarios. One could also consider fidelities for quenches in higher dimensions using dimensional reduction, similarly to the entanglement entropy. Moreover, it would be desirable to extend our results to the case of interacting quantum systems, integrable or not, though we anticipate that this is an extremely challenging direction. Finally, the RLE has a clear experimental relevance, and we are convinced that the various research ideas discussed above should also be investigated in realistic settings. In particular, a natural question is whether the nested lightcones structure of the RLE can be observed experimentally.  

\section*{Acknowledgements}

GP thanks Cl\'ement Berthiere and Rufus Boyack for useful discussions. 
This study was carried out within the National Centre on HPC, Big Data and Quantum Computing - SPOKE 10 (Quantum Computing) and received funding from the European Union Next- GenerationEU - National Recovery and Resilience Plan (NRRP) – MISSION 4 COMPONENT 2, INVESTMENT N. 1.4 – CUP N. I53C22000690001. This work has been supported by the project ``Artificially devised many-body quantum dynamics in low dimensions - ManyQLowD'' funded by the MIUR Progetti di Ricerca di Rilevante Interesse Nazionale (PRIN) Bando 2022 - grant 2022R35ZBF.

\providecommand{\href}[2]{#2}\begingroup\raggedright\endgroup

\end{document}